\documentclass[preprint,12pt]{elsarticle}
\biboptions{sort&compress}

\usepackage{adjustbox}
\usepackage{amssymb}
\usepackage{amsmath}
\usepackage{array}
\usepackage[english]{babel}
\usepackage{balance}
\usepackage{booktabs} 
\usepackage[format=plain,justification=justified,singlelinecheck=false,
font={stretch=1.125,small},labelfont=bf,labelsep=space]{caption}
\usepackage{charter}
\usepackage{colortbl}
\usepackage{color,soul}
\usepackage{droidsans}
\usepackage{epstopdf}
\usepackage{extsizes}
\usepackage{fancyhdr}
\usepackage{float}
\usepackage{fnpos}
\usepackage[T1]{fontenc}
\usepackage[left=1.5cm, right=1.5cm, top=1.785cm, bottom=2.0cm]{geometry}
\usepackage{graphicx} 
\usepackage[colorlinks=true, citecolor=blue]{hyperref}
\usepackage{lastpage}
\usepackage{lineno}
\usepackage{makecell}
\usepackage{mathptmx}
\usepackage{microtype}
\usepackage{multirow}
\usepackage[numbers]{natbib}
\usepackage[abs]{overpic} 
\usepackage{pdfcomment}
\usepackage{sectsty}
\usepackage{setspace}
\usepackage{soul}
\usepackage{subcaption}
\usepackage{tabularx}
\usepackage{times}  
\usepackage[compact]{titlesec}
\usepackage[textwidth=1.5cm]{todonotes}
\usepackage[normalem]{ulem}
\usepackage{siunitx}

\sethlcolor{yellow}

\soulregister\ref7
\soulregister\pageref7
\soulregister\NbC7
\soulregister\NbCO7

\journal{Additive Manufacturing}

\begin{document}

\begin{frontmatter}



\title{Laser Powder Bed Fusion Melt Pool Dynamics for Different Geometric Variations and Powder Layer Heights: High-Fidelity Multiphysics Modeling vs 2025 NIST Experiments}

\author[BUET]{Badhon Kumar\fnref{fn1}} 
\author[BUET]{Rakibul Islam Kanak}
\author[UD,UD-Physics]{Nishat Sultana}
\author[NU]{Jiachen Guo}
\author[UD]{Andrew Schrader}
\author[NU,HIDENN]{Wing Kam Liu}
\author[UD]{Abdullah Al Amin\fnref{fn1}}

\fntext[fn1]{$\dagger$ Corresponding authors. Emails: aamin1@udayton.edu}

\affiliation[BUET]{%
organization={Department of Mechanical Engineering},
institution={Bangladesh University of Engineering and Technology},
city={Dhaka},
postcode={1000},
country={Bangladesh}}

\affiliation[UD]
{%
organization={Department of Mechanical and Aerospace Engineering},
institution={University of Dayton},
addressline={300 College Park Drive},
city={Dayton},
state={OH},
postcode={45469},
country={USA}
}

\affiliation[UD-Physics]
{%
organization={Department of Physics},
institution={University of Dayton},
addressline={300 College Park Drive},
city={Dayton},
state={OH},
postcode={45469},
country={USA}
}

\affiliation[NU]
{%
organization={Department of Mechanical Engineering, Northwestern University},
addressline={2145 Sheridan Road, Room B224},
city={Evanston},
state={IL},
postcode={60208},
country={USA}
}

\affiliation[HIDENN]
{%
organization={HiDeNN-AI LLC},
addressline={1801 Maple Ave},
city={Evanston},
state={IL},
postcode={60201},
country={USA}
}

\begin{abstract}
Metal Laser Powder Bed Fusion (PBF-LB/M) is a leading additive manufacturing technique in which part quality and grain morphology are highly dependent on process parameters. Numerous studies of process variations, such as laser power, scan speed, and spot diameter, have demonstrated that they strongly influence melt pool dynamics; however, the effects of powder layer height and geometric variations remain less well understood. In this article, we focus on variations in powder layer height and part geometry to study their influence on melt pool dynamics. We employed a high-fidelity multiphysics simulation framework based on the open source finite volume method (FVM) solver package `LaserBeamFoam' built on `OpenFOAM' to study the variations in different melt pool metrics--- melt pool depth, width, bead height, overlap depth, overlap width, solidified area, and dilution area. The solver captures coupled phenomena of heat transfer, fluid flow, vaporization, recoil pressure, Marangoni convection, and realistic laser reflection behavior to accurately model the melt pool dynamics. Simulations are performed for different powder layer heights and geometric dimensions for direct comparison with benchmark experiments conducted at the National Institute of Standards and Technology (NIST) in 2025. Quantitative validation against NIST experiment demonstrates excellent agreement in all the melt pool metrics. These results highlight the predictive capability of physics-based PBF-LB models, paving the way for process optimization, defect mitigation, and the integration of simulation into digital twin frameworks for additive manufacturing.
\end{abstract}


\begin{keyword}
Laser Powder Bed Fusion \sep Metal Additive Manufacturing \sep AM Process Modeling



\end{keyword}

\end{frontmatter}

\section{Introduction}
\label{sec:intro}

Metal Laser Powder Bed Fusion (PBF-LB/M) has emerged as one of the most widely adopted additive manufacturing technologies for producing complex metallic components with high geometric precision and design flexibility \cite{zafarReviewMetalAdditive2025, joshuaPowderBedFusion2024, zhaoLaserPowderBed2023,herzogAdditiveManufacturingMetals2016, gorsseAdditiveManufacturingMetals2017}. Such components are widely used in fields such as aerospace engineering \cite{tepyloLaserBasedAdditiveManufacturing2019}, biomedical applications \cite{huangLaserPowderBed2020}, and advanced energy systems \cite{cramerAdditiveManufacturingCeramic2022}, where intricate geometries, lightweight structures, and high performance are required. It has demonstrated successful application in a wide range of material systems such as stainless steel \cite{ahmedProcessParameterSelection2022}, nickel alloys \cite{mostafaeiAdditiveManufacturingNickelbased2023}, titanium alloys \cite{eskandarisabziPowderBedFusion2019}, ceramics \cite{dadkhahAdditiveManufacturingCeramics2023}, etc. Although promising, PBF-LB/M faces significant challenges in terms of part quality and performance as the process experiences porosity \cite{mukherjeeMitigationLackFusion2018,liuPredictingPorosityDefects2022,bayatKeyholeinducedPorositiesLaserbased2019,weiMechanismsIntertrackVoid2020}, cracking \cite{vermaExtendedFiniteElement2022,yimCrackingBehaviorTi48Al2Cr2Nb2023}, residual stress induced failures \cite{bhatPredictionExperimentalVerification2022,levkulichEffectProcessParameters2019}, etc. Thus, it is important to fundamentally understand the process that includes complex laser-material interactions \cite{yeEnergyCouplingMechanisms2019,yinHighpowerLasermatterInteraction2019}, heat transfer \cite{aminPhysicsGuidedHeat2024,sarkarAdvancesComputationalModeling2024,ganNumericalSimulationThermal2017}, fluid flow \cite{aminPhysicsGuidedHeat2024,liuNumericalSimulationLaser2026,parivendhanNumericalStudyProcessing2023}, vaporization \cite{zhaoLaserMeltingModes2022}, recoil pressure \cite{luEffectMoltenPool2024}, and keyhole dynamics \cite{ganUniversalScalingLaws2021, wangMechanismKeyholePore2022,wangEffectsEnergyDensity2021}. Together, these phenomena govern melt pool morphology and strongly influence the resulting microstructure \cite{solaMicrostructuralPorosityAdditive2019}, defect formation \cite{kyogokuReviewMetalAdditive2020}, and overall part quality \cite{chenReviewQualificationCertification2022}. In addition, poor melt pool dynamics can lead to defects such as lack-of-fusion \cite{mojumderLinkingProcessParameters2023}, balling \cite{liBallingBehaviorStainless2012}, and hot cracking \cite{chenSituObservationReduction2024} which directly affect the structural integrity of additively manufactured components.

Over the past decade, significant research efforts have focused on developing physics-based computational models to understand and predict melt pool behavior in PBF-LB/M processes \cite{liReviewMetalAdditive2019,afrasiabiModellingSimulationMetal2023, bouabbouUnderstandingLasermetalInteraction2022,xuNumericalSimulationMelt2022, nabaviComprehensiveReviewRecent2024, akterjahanMultiscaleModelingFramework2022, weiMechanisticModelsAdditive2021}. Continuum-scale multiphysics models based on computational fluid dynamics (CFD) leveraging methods such as Finite Difference Methods (FDM) \cite{renFiniteDifferenceMethod2023}, Finite Volume Methods (FVM) \cite{aminPhysicsGuidedHeat2024}, Finite Element Methods (FEM) \cite{leonorGOMELTGPUoptimizedMultilevel2024,trejos-tabordaFiniteElementModeling2022}, Lattice Boltzmann Method (LBM) \cite{zakirovPredictiveModelingLaser2020}, and Smoothed Particle Hydrodynamics (SPH) \cite{russellNumericalSimulationLaser2018, gingoldSmoothedParticleHydrodynamics1977} have been widely used to capture the coupled thermo-fluid phenomena governing melt pool formation. In addition, reduced order methods such as Proper Orthogonal Decomposition (POD) \cite{luAdaptiveHyperReduction2020} and Proper Generalized Decomposition (PGD) \cite{stroblPGDThermalTransient2024} also attempted to solve and study melt pool behavior in an accelerated fashion. These models typically solve conservation equations for mass, momentum, and energy. Additional physical mechanisms such as Marangoni convection, buoyancy forces, surface tension, recoil pressure from evaporation, and laser energy absorption are also incorporated. Using these approaches, researchers have been able to reproduce keyhole formation, melt pool geometry, and solidification and cooling behavior with increasing accuracy \cite{aminPhysicsGuidedHeat2024, ganBenchmarkStudyThermal2019}. In the interest of computational efficiency, lately there have been a push in data driven numerical methods such as physics informed neural networks (PINNs) \cite{zhangScientificDeepLearning2026}, surrogate modeling \cite{luAdaptiveHyperReduction2020}, and innovative novel solution approaches \cite{guoTensordecompositionbasedPrioriSurrogate2025,zhangMultiLevelVariationalMultiScale2026}. These methods leverage data from high-fidelity simulations or experiments to train models that can rapidly predict melt pool characteristics under varying processing conditions and in some cases converge to a solution without any datasets.

Despite these advances, accurately predicting melt pool characteristics under different processing conditions remains challenging. In particular, accurately including the effects of different process parameters such as laser power, scan speed, spot diameter, scan hatch spacing, powder layer thickness, geometric configurations and scanning strategy parameters in the model has been challenging. The simplest of the modeling approach is to consider the heat diffusion equation \cite{masoomiNumericalExperimentalInvestigation2018} which is relatively easy to solve by ignore some of the key physical descriptions. Improvement on top of the simple heat diffusion equation would be to introduce flow behavior of the molten metals, namely consider the Navier Stokes equation \cite{aminPhysicsGuidedHeat2024,ganBenchmarkStudyThermal2019}. These models capture some of the melt pool behavior and can predict the melt pool depth and width along with cooling rates at a reasonable accuracy. However, to be able to predict the bead height, and melt pool depth and width overlap, surface evolution needs to be captured in the model. Different techniques such as Volume of Fluid (VOF) \cite{hirtVolumeFluidVOF1981}, Level-set (LS) \cite{dervieuxFiniteElementMethod1980}, multiphase modeling \cite{chenSpatteringDenudationLaser2020} and Smoothed Particle Hydrodynamics (SPH) \cite{gingoldSmoothedParticleHydrodynamics1977} needs to be introduced. Next, the accurate heat source model further improves the heating and cooling behavior of the model which is necessary for the most accurate melt pool morphology under different process variations. There have been a multitude of heat source models--- starting with the traditional Goldack heat source \cite{goldakNewFiniteElement1984} to surface \cite{leeModelingHeatTransfer2016} and volumetric heat source \cite{zhang3DimensionalHeatTransfer2019} to the most accurate Ray-tracing \cite{liuNewRayTracing2020} based heat source. Ray tracing algorithm generally better predict the resulting keyhole during the LPBF process \cite{renHighfidelityModellingSelective2021, leStudyKeyholemodeMelting2019} but the approch becomes increasingly computationally expensive. Leveraing all these detailed physical discription, the computational model of PBF-LB/M process can be utilized to study melt pool morphologies for different process conditions \cite{aminPhysicsGuidedHeat2024, ganBenchmarkStudyThermal2019, zhangDefectsCausedPowder2025}. However, these efforts are largely limited to single-track simulations as far as the computational cost is of concern. Fewer studies have investigated the combined effects of powder bed characteristics and multi-track scanning in a unified modeling framework that considers all the key physics. In addition, experimental validation of high-fidelity simulations is still limited because of the scarcity of the well-controlled benchmark datasets. But it is important to understand the melt pool dynamics under different powder layer heights and geometric variations as these parameters determines the part level performance. 

In recent years, the National Institute of Standards and Technology (NIST) introduced the Additive Manufacturing Benchmark (AM-Bench) challenges (starting in 2018) to understand the AM process with greater details and thus provide support for the validation of the computational models. These initiatives provide carefully controlled experimental datasets. One of the primary goal of these benchmark tests is to allow researchers to test their predictive simulations against highly controlled experimental data, contributing to the broader effort of developing predictive modeling and simulation tools crucial for the qualification and certification of additively manufactured components. One of the NIST experiment specifically focuses on the effect of varying powder layer thickness on the of part size of different dimensions. The experiments for this challenge were performed using the NIST Fundamentals of Laser-Material Interaction (FLaMI) metrology platform. Specifically titled AMB2025-06-PMPG challenge, explained in the NIST challenge 06 description document \cite{nist_amb2025_results}, the experiment examines three distinct powder layer conditions on two different part dimensions: $5 \,\text{mm} \times 5 \,\text{mm}$ and $1 \,\text{mm} \times 5 \,\text{mm}$. The experimental conditions comprise of a bare substrate ($0 \,\mu\text{m}$ powder), an $80 \,\mu\text{m}$ powder layer, and a $160 \,\mu\text{m}$ powder layer. Process parameters remain constant across all tests: laser power of $285 \,\text{W}$, scanning speed of $960 \,\text{mm/s}$, hatch spacing of $0.11 \,\text{mm}$, and Gaussian beam spot diameter of $72 \,\mu\text{m}$. The challenge measurs melt pool metrics through in-situ measurements of temperature measurements and ex-situ characterization of cross-section micrographs.

The core modeling tasks for AMB2025-06 are divided into two challenge problems. The first is the Pad Melt Pool Geometry (CHAL-AMB2025-06-PMPG), which requires modelers to calculate geometric measurements for each pad cross-section, specifically the average bead height, depth, overlap depth, and width. Additionally, modelers were challenged to calculate the total solidified area above the substrate and the total dilution area below the substrate. The second problem is the Pad Surface Topography (CHAL-AMB2025-06-PST), which requires the calculation of the fused layer thickness and the root-mean-square-height ($S_q$). The required surface topography metrics must be calculated for various specified evaluation areas and profiles within the $5 \,\text{mm} \times 5 \,\text{mm}$ and $1 \,\text{mm} \times 5 \,\text{mm}$ pad geometries. Building on these benchmark challenges and available datasets, this study develops a high-fidelity multiphysics simulation framework to investigate laser-material interactions of the melt pool morphology in the PBF-LB/M process. The model is implemented using the open-source finite volume solver `LaserBeamFoam' \cite{flintLaserbeamFoamLaserRaytracing2023, flintVersion20LaserbeamFoam2024, flintVersion30LaserbeamFoam2025}. It captures key physical phenomena in the melt pool, including heat transfer, fluid flow, vaporization, recoil pressure, and free-surface evolution using a Volume of Fluid (VOF) formulation. Laser energy absorption is modeled using a Gaussian heat source together with Fresnel-based absorptivity and a ray-tracing approach that accounts for multiple reflections inside the evolving keyhole cavity. To generate powder-bed particles, the open-source DEM solver \textit{LIGGGHTS}~\cite{LIGGGHTSOpenSource} is used. Past numerical studies typically examined single tracks or, in limited cases, up to 8 tracks \cite{quevaMesoscaleMultilayerMultitrack2024} with limited physical descriptions, however, this challenge required results averaged from 45-track simulations, thereby challenging the current state-of-the-art computational modeling capabilities. In this work, simulations are performed for different powder layer thicknesses and part dimensions as a direct representation to the NIST AM-Bench 2025 benchmark experiments that include part-scale simulations of 45 parallel laser scan tracks. The numerical predictions are quantitatively compared with experimental measurements of melt pool depth, width, bead height, and overlap characteristics and presented as a validation case.

\begin{figure}
    \centering
    \includegraphics[width=0.8\textwidth]{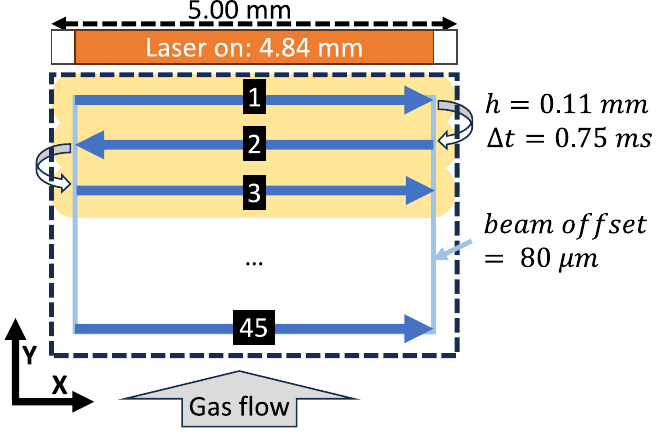}
    \caption{Illustration of laser scan strategy for 5 mm x 5 mm pad with a 0.75 ms laser turnaround time. The image is taken from NIST AM Bench 2025 Challenge description \cite{nist_amb2025_results}}
    \label{fig:scan_strategy}
\end{figure}

\begin{figure}
    \centering
    \includegraphics[width=\textwidth]{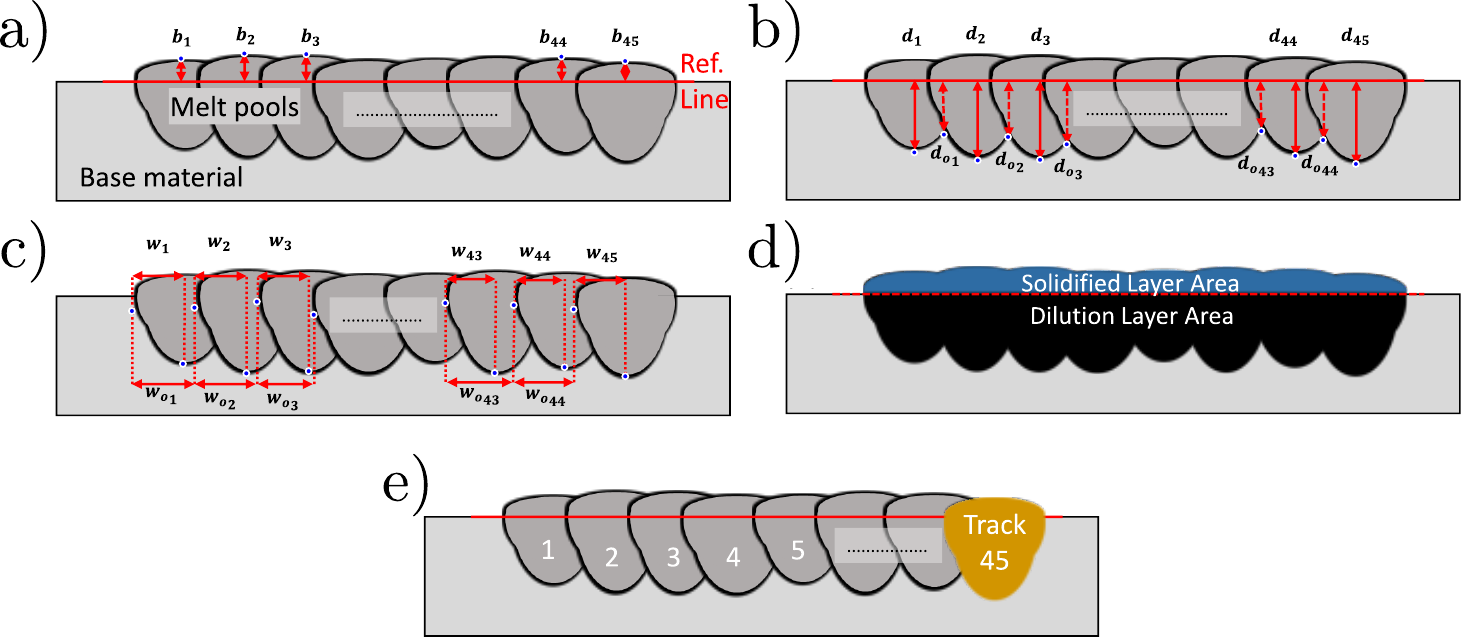}
    \caption{Cross-sectional schematic of melt-pool measurands; (a) bead height, (b) depth and overlap depth, (c) width and overlap width, (d) solidified and dilution layer areas, and (e) representative melt-pool image of the final track. Images are taken from NIST AM Bench 2025 Challenge description \cite{nist_amb2025_results}}
    \label{fig:schematic_meltpool}
\end{figure}

\section{Methodology}
\label{sec:method}

The laser beam heating process was modeled using LaserbeamFoam, an open-source thermo-fluid solver developed within the OpenFOAM-10 C++ Finite Volume framework \cite{flintLaserbeamFoamLaserRaytracing2023}. This solver employs the multiphase Volume of Fluid (VOF) method to accurately capture the sharp interface between the metallic substrate and the surrounding gaseous domain. A ray-tracing algorithm based on Fresnel absorption equations is used to model the laser-material interaction with high fidelity.

The model incorporates several key assumptions to balance computational efficiency with physical accuracy. All phases (solid, liquid, and gas) are treated as incompressible Newtonian fluids under laminar flow conditions. Density variations arising from temperature changes are accounted for exclusively in the gravity term ($\mathbf{F}_{g}$) through the Boussinesq approximation, maintaining consistency with the incompressibility assumption. The complex physics of metal vaporization and vapor plume dynamics are simplified by applying recoil pressure ($P_v$) as a surface force at the liquid-gas interface, rather than explicitly modeling volumetric expansion. The metallic substrate is assumed opaque, with all laser energy absorption and reflection occurring at the VOF interface. Material electrical resistivity is considered temperature-independent for computational simplicity. The laser beam is modeled as unpolarized, with total absorptivity calculated by averaging the reflectance of perpendicularly and parallelly polarized light according to Fresnel equations. Finally, given the fiber laser wavelength of 1.064~$\mu$m employed in this study, the inverse Bremsstrahlung effect is neglected \cite{katayamaElucidationLaserWelding2010}.

To predict the transient temperature and velocity fields in the melt pool region, three sets of conservation equations - mass, momentum and energy equations are solved.

The mass conservation equation is expressed as,

\begin{equation}
\nabla \cdot \mathbf{U} = 0
\label{eq:mass}
\end{equation}

where $\mathbf{U}$ is the velocity. The momentum conservation equation is expressed as follows, 

\begin{equation}
\frac{\partial(\rho \mathbf{U})}{\partial t} 
+ \nabla\cdot(\rho \mathbf{U}\otimes \mathbf{U}) 
= -\nabla P + \nabla\cdot\boldsymbol{\tau} 
+ \mathbf{F}_{g} + \mathbf{F}_{st} 
+ \mathbf{F}_{\text{mush}} + \mathbf{F}_{\text{rec}}
\label{eq:momentum}
\end{equation}

Where, $\rho$ is the mass density, $P$ is the fluid pressure, and $\boldsymbol{\tau}$ is the viscous stress tensor. In Eq. \ref{eq:momentum}, additional volumetric source terms are incorporated to account for relevant physics that govern the motion of the molten metal. The buoyancy force $\mathbf{F}_{g}$ is calculated using the Boussinesq approximation. 

\begin{equation}
\mathbf{F}_g = \rho \mathbf{g} \beta (T - T_{ref})
\end{equation}

where $\beta$ is the thermal expansion coefficient and $T_{\text{ref}}$ is the reference temperature. The surface tension force and Marangoni convection are accounted for by the term $\mathbf{F}_{st}$, which is expressed as,

\begin{equation}
    \mathbf{F}_{st} = \left( \sigma \kappa \mathbf{n} + \frac{d\sigma}{dT} \left[\nabla T - \mathbf{n}(\mathbf{n} \cdot \nabla T)\right] \right) |\nabla \varphi_m| \frac{2\rho}{\rho_m + \rho_g}
\end{equation}

where, $\sigma$ is the surface tension coefficient, $\kappa$ is the surface curvature of the metal free surface, $\mathbf{n}$ is the unit surface normal vector, $|\nabla \varphi_m|$ is the magnitude of the metal volume fraction, which converts surface forces to volumetric force and the multiplier term $\frac{2\rho}{\rho_m + \rho_g}$ redistributes the interfacial forces towards the heavier phase (metal) with $\rho_m$ and $\rho_g$ as density of metal and gas respectively. $\frac{d\sigma}{dT}$ denotes the temperature gradient of surface tension.

The solidification and melting processes are captured using the enthalpy-porosity technique \cite{brentENTHALPYPOROSITYTECHNIQUEMODELING1988}. This approach introduces a volumetric damping source term, commonly referred to as the Carman-Kozeny sink term, which appears as $\mathbf{F}_{\text{mush}}$ in the momentum equation (Eq. \ref{eq:momentum}). The damping force is expressed as

\begin{equation}
    \mathbf{F}_{\text{mush}} = \frac{(1-\epsilon)^2}{(\epsilon^3+ 10^{-12})} A_{\text{mush}} (\mathbf{U}-\mathbf{U}_{p})
\end{equation}

Here, $\epsilon$ is the cell liquid fraction, equal to $1$ in solid cells, $0$ in liquid cells, and between $0$ and $1$ in the mushy zone. $A_{\text{mush}}$ is the mushy zone constant, which is taken as $10^{6}\ \mathrm{kg\,m^{-3}\,s^{-1}}$ \cite{alphonsoPossibilityDoingReduced2023}. $\mathbf{U}_{p}$ is the solid velocity due to the pulling of solidified material out of the domain.

The variable $\epsilon$ represents the volume fraction available for flow or permeability. The definition of $\epsilon$ is divided into two distinct modes to capture the physical differences between a solid substrate and powder bed.

In the default configuration no powder particles are considered, which is similar to a solid substrate model. The Carman-Kozeny sink term $\mathbf{F}_{\text{damp}}$ is calculated using thermodynamic liquid fraction $\epsilon$.

\begin{equation}
\epsilon = 
\begin{cases} 
0 & \text{if } T < T_S \\ 
(T-T_S)/(T_L-T_S) & \text{if } T_S \leq T \leq T_L \\
1 & \text{if } T > T_L 
\end{cases}
\end{equation}

However, a masked liquid fraction is introduced in the model denoted as $\epsilon_\text{mask}$ which replaces the standard $\epsilon$ where powder particles are present. This variable enforces a rigid state until the powder particles are fully molten. This masked liquid fraction serves as a numerical switch that distinguishes between the coherent melt pool and the loose powder. It is derived from the standard temperature-dependent liquid fraction $\epsilon$ using a threshold filter:

\begin{equation}
\epsilon_{\text{mask}} = 
\begin{cases} 
0 & \text{if } \epsilon \leq 0.95 \\ 
\epsilon & \text{if } \epsilon > 0.95 
\end{cases}
\end{equation}

Physically, this ensures that the drag coefficient in the Carman-Kozeny sink term remains at its maximum value, effectively infinite drag, even when the powder is partially heated. Consequently, the velocity $\mathbf{u}$ is forced to zero in the powder regions, effectively freezing the powder particles in space until they are fully encompassed by the melt pool ($\epsilon > 0.95$). This treats the semi-solid region as an impermeable, fully dense barrier rather than a permeable porous medium.

Intense volumetric laser heating induces material vaporization, generating a surface force on the molten metal known as recoil pressure. This pressure creates a surface depression termed a keyhole. Under atmospheric conditions, the recoil pressure typically ranges from 0.5 to 0.6 times the saturated vapor pressure at the vaporization temperature (at 1 atm) due to recondensation effects following evaporation \cite{wangMechanismKeyholePore2022}.

The volumetric recoil pressure source, $\mathbf{F}_{rec}$ is expressed as, 
\begin{equation}
       \mathbf{F}_{rec} = (P_r - P_{amb})\mathbf{n}|\nabla \varphi_m| \frac{2\rho}{\rho_m + \rho_g}
\end{equation} 

where, $P_r$ and $P_{amb}$ are the recoil pressure and ambient pressure respectively. The recoil pressure $P_{r}$ can be expressed as \cite{liNumericalExperimentalStudy2019}, 

\begin{equation}
P_r =
\begin{cases} 
    \dfrac{1+\beta_R}{2} P_{sat}, & T \geq T_b \\[6pt]
    P_{\text{amb}}, & 0 \leq T < T_b
\end{cases}
\end{equation}
where, $\beta_R$ is the retro-diffusion coefficient, which depends on the Mach number at the exit of the Knudsen layer. Its value changes from 0.18 to 1, with decrease in ambient pressure \cite{knightEvaporationCylindricalSurface1976}. $T_b$ is the evaporating temperature at corresponding ambient pressure. $P_{sat}$ is the temperature dependent vapor pressure and is calculated using the Clausius-Clapeyron law \cite{linModernThermodynamicsHeat1999}, which is expressed as, 
\begin{equation}
P_{\text{sat}} = P_{\text{amb}} \, \exp\!\left( \frac{M_l L_{v}}{R} \, \frac{T - T_{b}}{T \, T_b} \right)
\end{equation}

where, $M_l$ is the molar mass of the vaporized metal, R is the gas constant, and $L_v$ is the latent heat of vaporization.

Finally, Eq.~\ref{eq:energy} describes the conservation of energy in the computational domain:

\begin{equation}
\frac{\partial(\rho c_{p}T)}{\partial t} 
+ \nabla \cdot (\mathbf{U}\rho c_{p}T) 
- \nabla \cdot (k \nabla T) 
= Q_{\text{laser}} + S_{h} - Q_{\text{losses}}
\label{eq:energy}
\end{equation}

where, $C_p$ and $k$ are the specific heat and thermal conductivity of the mixture of two phases (gas and molten metal). In our work, they are temperature dependent. The phase change effects during the melting are calculated using the $S_h$ term. 

The term $Q_{losses}$ in the energy conservation equation (Eq. \eqref{eq:energy}) represents the heat losses during the laser welding process, which includes the convective ($Q_{conv}$), radiative ($Q_{rad}$) and Evaporative ($Q_{evap}$) heat loss terms defined with, 

\begin{align}
Q_{\text{losses}} &= \bigl( Q_{\text{conv}} + Q_{\text{rad}} + Q_{\text{evp}} \bigr) 
\left|\nabla \varphi_m\right| 
\left( \frac{2\rho}{\rho_m + \rho_g} \right) \label{eq:Qlosses} \\
Q_{\text{rad}}   &= \sigma \, \varepsilon \left( T^4 - T_{\infty}^4 \right) \\
Q_{\text{conv}}  &= h \left( T - T_{\infty} \right) \\
Q_{\text{evp}}   &= \dot{m}L_v 
\end{align}

Here, $\sigma$ is the Stephan-Boltzman constant, $\epsilon$ is emissivity of the material, $h$ is convective heat transfer coefficient, $T_\infty$ is the ambient temperature. $\dot{m}$ is the mass transfer rate during evaporation, and is approximated using the Hertz-Langmuir relation \cite{kolasinskiSurfaceScienceFoundations2012}, which is written as, 

\begin{equation}
\dot{m} =
\begin{cases}
0, & T < T_b \\[2mm]
(1-\beta_R) \, \sqrt{\dfrac{M_l}{2 \pi R T}} \, P_{\text{sat}}, & T \ge T_b
\end{cases}
\end{equation}

The laser heat source model ($Q_{laser}$) in Eq.~\ref{eq:energy} follows a Gaussian distribution that scans through the top surface of the substrate at a constant speed and can be expressed as, 

\begin{equation}
    Q_{laser} = I_0(r)(\mathbf{I_0 \cdot n_0})\eta(\theta_0) + \sum_{m=1}^{N} I_m(r)(\mathbf{I_m \cdot n_m})\eta(\theta_m)
\end{equation} 

with the incident laser intensity $I_0(r)$ defined as,

\begin{equation}
I_0(r) = \frac{2P}{\pi r_0^2 \Delta} \, \exp\!\left[ \frac{-2 (r - V_s t)^2}{r_0^2} \right]
\end{equation}

here, $\theta$ is the angle between the incident laser beam and normal vector of the molten metal interface, $\eta$ is the Fresnel absorption coefficient, N is the number of laser beam incidences considering multiple reflections, $\mathbf{I}$ is normalized laser beam direction, $\mathbf{n}$ is normalized normal vector of the molten material interface, and $I$ represents the laser energy intensity. The subscript $0$ refers to the incident beam and $m$ denotes the $m^{th}$ reflection. $P$ is the total deposited beam power, $\Delta$ is the computational cell length. $r_0$ is the beam radius($1/e^2$) and $V_s$ is the laser scanning speed. 

To account for multiple reflection during the laser welding process, a high-fidelity ray-tracing method has been implemented in the solver. When each ray hits a computational cell within the domain, part of it gets absorbed and the rest is reflected. The reflection vector is computed using the following equation, 

\begin{equation}
    \mathbf{V} _R = \mathbf{V}_I - \frac{2V_I \cdot \mathbf{n}}{|\mathbf{n^2}|} \mathbf{n}
\end{equation}

where $\mathbf{V}_R$ is the reflected ray vector and $\mathbf{V}_I$ is the incident ray vector.

The laser absorptivity $\eta$ of the incoming ray energy is calculated by the Fresnel equations \cite{Cho_2006, hanStudyLaserKeyhole2021} and is given by, 

\begin{equation}
    \eta = 1 - \frac{R_s +R_p}{2}
\end{equation}

where, $R_s$ and $R_p$ are the reflectance for parallel and perpendicularly polarised light defined as:
\begin{align}
R_{s} &= \frac{\chi^{2} + \psi^{2} - 2 \chi \cos\theta + \cos^{2}\theta}
             {\chi^{2} + \psi^{2} + 2 \chi \cos\theta + \cos^{2}\theta} \label{eq:Rs} \\
R_{p} &= R_{s} \, 
        \frac{\chi^{2} + \psi^{2} - 2 \chi \sin\theta \tan\theta + \sin^{2}\theta \tan^{2}\theta}
             {\chi^{2} + \psi^{2} + 2 \chi \sin\theta \tan\theta + \sin^{2}\theta \tan^{2}\theta} \label{eq:Rp}
\end{align}

where, $\theta$ is the incidence angle between a particular discretised ray and the substrate where the ray interacts. $\chi$ and $\psi$ are the functions of the refractive index, which can be written as, 
\begin{align}
\chi &= \left( \frac{\sqrt{(n^{2} - k^{2} - \sin^{2}\theta)^{2} + 4 n^{2} k^{2}} + n^{2} - k^{2} - \sin^{2}\theta}{2} \right)^{\frac{1}{2}} \label{eq:chi} \\
\psi &= \left( \frac{\sqrt{(n^{2} - k^{2} - \sin^{2}\theta)^{2} + 4 n^{2} k^{2}} - n^{2} + k^{2} + \sin^{2}\theta}{2} \right)^{\frac{1}{2}} \label{eq:psi}
\end{align}

Here, $n$ is the refractive index and $k$ is the extinction coefficient. The expressions of $n$ and $k$ are as follows: 
\begin{align}
n &= \left( \frac{\sqrt{e_{r}^{2} + e_{i}^{2}} + e_{r}}{2} \right)^{\frac{1}{2}} \label{eq:n} \\
k &= \left( \frac{\sqrt{e_{r}^{2} + e_{i}^{2}} - e_{r}}{2} \right)^{\frac{1}{2}} \label{eq:k}
\end{align}

The terms $e_r$ and $e_i$ in Eq.~\ref{eq:n} and Eq.~\ref{eq:k} are the real and imaginary parts of the relative electric permittivity, respectively, which are defined as:
\begin{align}
e_{r} &= 1 - \frac{\omega_{p}^{2}}{f^{2} + \delta^{2}} \label{eq:er} \\
e_{i} &= \frac{\delta}{f} \, \frac{\omega_{p}^{2}}{f^{2} + \delta^{2}} \label{eq:ei}
\end{align}

Here, $\omega_p$, $f$, and $\delta$ are the plasma frequency, incident laser angular frequency and damping frequency, respectively which are given by, 
\begin{align}
\omega_{p} &= \sqrt{\frac{N_{e} q_{e}^{2}}{M_{e} \epsilon_{0}}} \label{eq:omega_p} \\
f &= \frac{2 \pi c}{\lambda} \label{eq:f} \\
\delta &= \omega_{p}^{2} \, \epsilon_{0} R_{e} \label{eq:delta}
\end{align}

Here, $N_e$ is the mean electron density of the substrate material, $q_e$ and $M_e$ are the charge and mass of an electron respectively, $\epsilon_0$  is the vaccum permittivity, $c$ is the speed of light, $\lambda$ is the laser wavelength and $R_e$ is the electric resistivity of the substrate material. $N_e$ and $R_e$ have been used to calibrate the computational model.

To capture the gas-liquid interface, the Volume of Fluid (VOF) \cite{leeModelingHeatTransfer2016} method is used. The VOF equation can be written as, 

\begin{equation}
\frac{\partial \alpha}{\partial t} + \nabla \cdot (\mathbf{U} \alpha) = 0
\end{equation}

where, $\alpha$ implies the volume fraction of the metallic phase within the cell. $\alpha$ = 1 implies, the cell is filled with metal phase, whereas $\alpha$ = 0 means the cell is entirely filled with gas phase. Values in-between indicates of the existance of free surface.

\section{Solver setup \& Model Validation}
\label{sec:Validation}

The computational domain is a rectangular box of dimensions
5.14~mm $\times$ 1.94~mm $\times$ 0.6~mm (length $\times$ width $\times$ height) for a 5mm $\times$ 5mm bed size and 1.84~mm $\times$ 1.14~mm $\times$ 0.6~mm (length $\times$ width $\times$ height) for a 1mm $\times$ 5mm bed size,
discretized with a uniform mesh spacing of 10~$\mu$m. A fixed time step of
$1 \times 10^{-7}$~s was employed, and the Courant number was held below 0.25
throughout all simulations to guarantee numerical stability. This spatial and
temporal resolution constitutes a continuum-scale discretization capable of
resolving the macroscopic features of the melt pool geometry,
in line with established practices in three-dimensional multi-physics laser
processing simulations
\cite{wangEffectsEnergyDensity2021, ganBenchmarkStudyThermal2019}.

\begin{figure}[!h]
    \centering
    \includegraphics[width=0.9\textwidth]{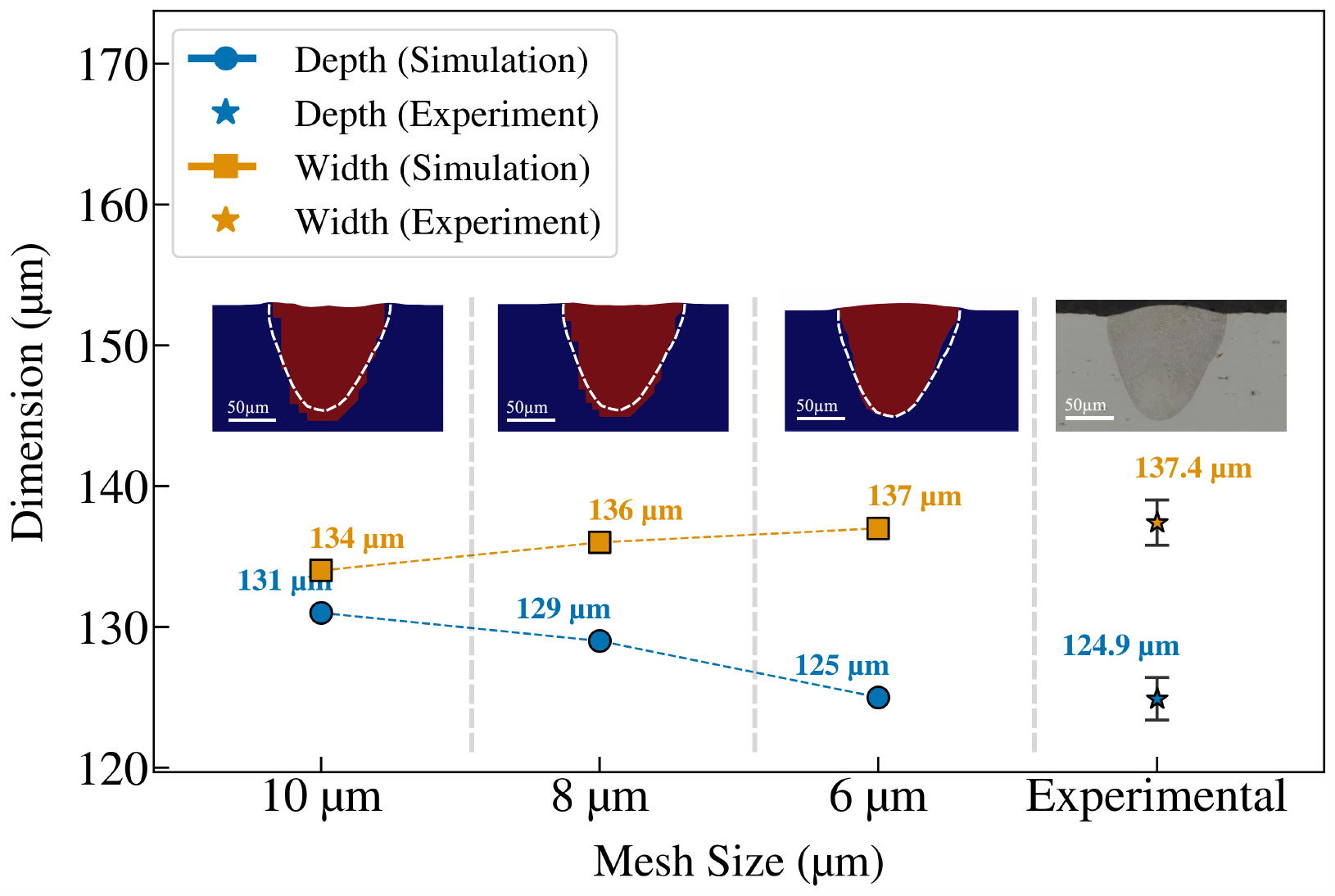}
    \caption{Grid sensitivity analysis of predicted melt pool depth and width at three mesh resolutions (10, 8, and 6~$\mu$m) compared against experimental measurements from the 2025 NIST AM-Bench Challenge~06 calibration dataset \cite{nist_amb2025_results}. White dashed lines drawn on the simulated figures represent the experimental melt pool shape. This figure is reproduced from an earlier work by one of the authors employing the same CFD solver \cite{kanakProcessMicrostructureCoupling2025}.}
    \label{fig:mesh_sensitivity}
\end{figure}

To assess the adequacy of the chosen mesh, a grid sensitivity study was carried
out using the `LaserbeamFoam' solver against the 2025 NIST AM-Bench
Challenge~06 dataset for single track laser scan \cite{nist_amb2025_results}. A single-track scan
on IN718 was simulated with a laser power of $P = 285$~W, a scan speed of
$v = 0.96$~m/s, and a laser spot diameter of $72\,\mu$m . Three mesh resolutions - $6\,\mu$m, $8\,\mu$m, and $10\,\mu$m were evaluated, and the predicted melt
pool depth and width were compared against experimental measurements, as shown in
Fig. \ref{fig:mesh_sensitivity}. Across all three resolutions the predictions
fall within a $5\%$ deviation from the experimental reference, confirming that the
$10\,\mu$m mesh delivers adequate predictive accuracy and is therefore adopted for 
the full-scale simulations presented in this work.

The solid-liquid phase transition is handled through the enthalpy-porosity
formulation, in which the liquid fraction field governs latent heat release and a
Darcy-type momentum sink damps velocities progressively across the mushy zone
\cite{dalINVITEDOverviewState2016,xiongEvaluateEffectMelt2022},
eliminating the need for nanometer-scale resolution at the phase boundary.
Interfacial forces including Marangoni stress, surface tension, and laser-induced
recoil pressure are incorporated as volumetric body forces through the Continuum
Surface Force (CSF) framework \cite{huSelectiveLaserMelting2021,keModelingNumericalStudy2021}, which projects these surface-localized effects onto the neighboring fluid cells and drives the thermocapillary convection that governs the final weld pool dimensions. This continuum modeling strategy has been thoroughly validated against experimental melt pool geometries and keyhole morphologies in a wide range of prior studies \cite{wangEffectsEnergyDensity2021, aggarwalInvestigationTransientCoupling2023}.

The thermo-fluid model was also calibrated against experimentally measured melt-pool dimensions for single track laser scan to ensure reliability under unified conditions, encompassing both bare-plate and varying powder-bed thickness configurations.

\begin{figure}[htbp]
    \centering
    \includegraphics[width=0.9\textwidth]{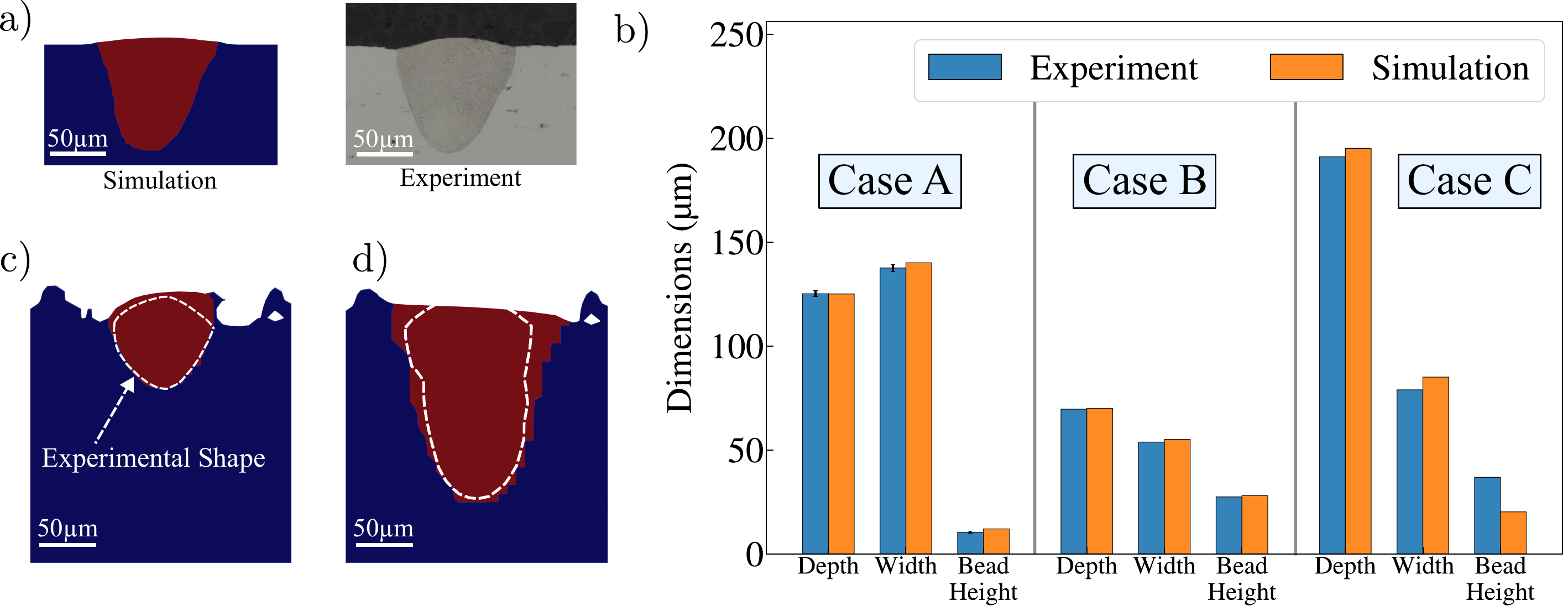}    
    \caption{Model calibration and validation against experimental meltpool data for IN718. \textbf{a} Qualitative comparison of simulated (left) and experimental (right) meltpool cross-sections for NIST AM Bench 2025 bareplate case ($P = 285$~W, $v = 0.96$~m/s, $d = 72$~$\mu$m). \textbf{b} Quantitative comparison of meltpool depth, width, and bead height for bareplate (Case A) and powder bed configurations (Case B : $P = 100 W$, $V_s = 1.0$ m/s, $d = 100 \mu$m, layer height = 50 $\mu$m, Case C : $P = 370 W$, $V_s = 0.8$ m/s, $d = 100 \mu$m, layer height = 50 $\mu$m). \textbf{c, d} Qualitative morphology comparison with experimental contours (dashed lines) overlaid on simulations for Cases B and C. Red regions indicate heat-affected zones (molten/solidified); blue regions remain solid.}
    \label{fig:model_validation} 
\end{figure}

At first, the validation was performed using single-track laser scans on bare IN718 plates, using experimental data from the AMB2025-06 benchmark challenge description data \cite{nist_amb2025_results}. Fig. \ref{fig:model_validation} dictates that the simulation results for melt pool width, depth, and bead height have strong agreement with the bare-plate measurements.

Second, to validate the model's performance for laser powder bed fusion (PBF), experimental results from Khorasani et al. \cite{khorasaniComprehensiveStudyMeltpool2022} were used as a benchmark. Two distinct processing conditions from this study were selected to test the model's reliability. For Case B and C, laser power was set to 100 W and 370 W respectively with laser scan speed of 1 m/s and 0.8m/s respectively.

The experimental measurements were extracted through digital image analysis of published metallographic cross-sections. Scale bars provided in the original images were used for calibration and the dimensions of the melt-pool were measured manually. Although this approach may introduce some uncertainty in the measurement, it allows quantitative comparison between simulation predictions and experimental observations.

At a lower energy density, represented by Case B as presented in Fig. \ref{fig:model_validation}b, the simulation demonstrates excellent predictive accuracy. The calculated melt pool depth, bead height and width are predicted with high fidelity, showing minimal errors within around 2\% error margin. At a higher energy density, represented by Case~C, the model again shows excellent agreement with the simulated depth of 195 $\mu$m at 2.09\% error from the experimentally measured depth of 191 $\mu$m, and the width shows a reasonable 7.73\% difference. However, the bead height prediction of 20.2 $\mu m$ deviated a little more from the experimental value of 36.8 $\mu $m.

\section{Results}

\begin{figure}[!h]  
    \centering
    \includegraphics[width=\textwidth]{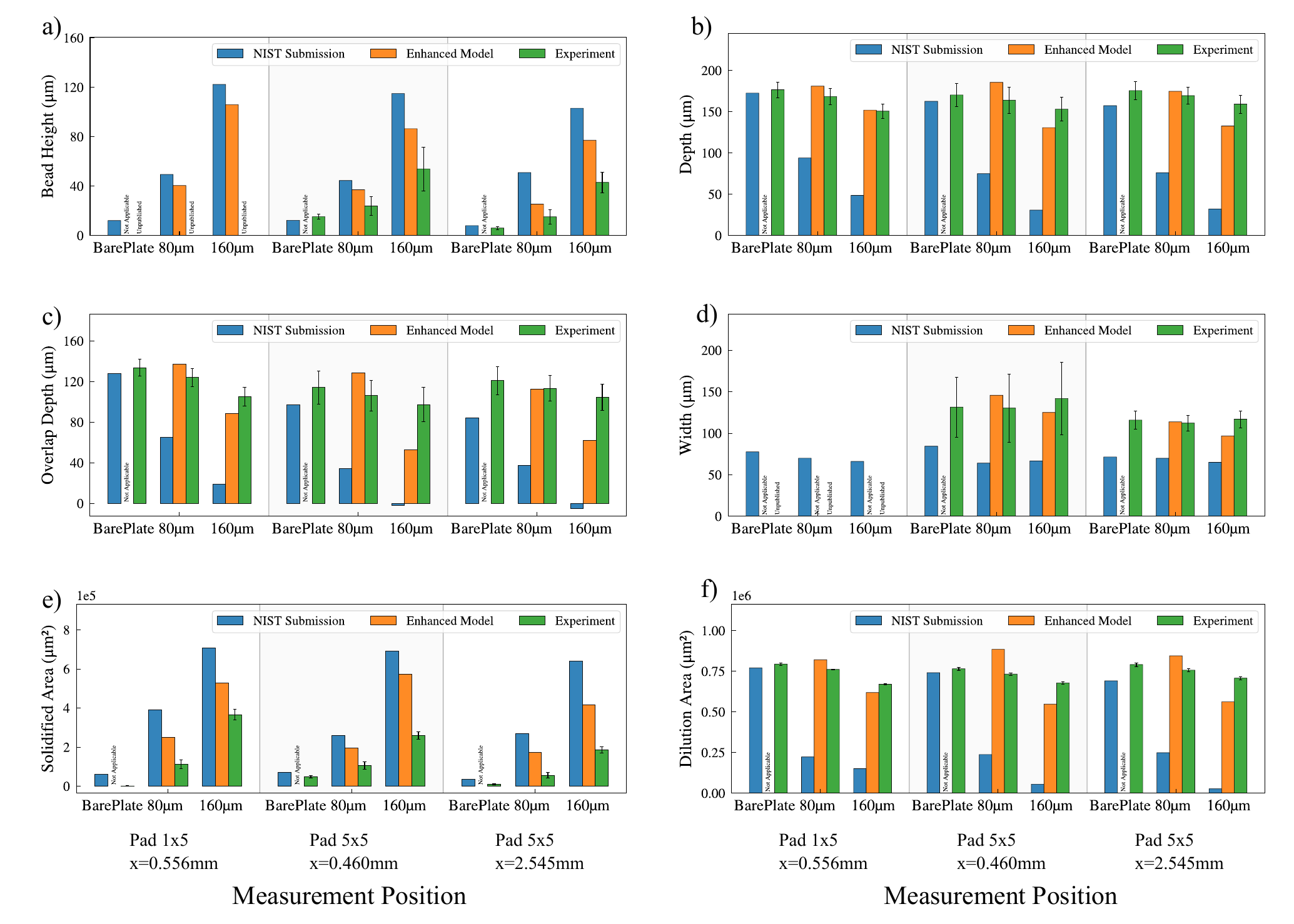}
    \caption{Quantitative validation of melt-pool geometric characteristics under varying process conditions. Comparison of (a) bead height, (b) melt-pool depth, (c) overlap depth, (d) melt-pool width, (e) total solidified area, and (f) total dilution area across three powder layer configurations: bare plate ($0\,\mu\text{m}$), $80\,\mu\text{m}$, and $160\,\mu\text{m}$. The results compare the initial NIST benchmark submission (blue) and the enhanced multiphysics model (orange) against experimental measurements (green). Results are shown for two pad geometries: a 1 mm $\times$ 5 mm pad (denoted as Case $1\times5$ in the figure, measured at $x = 0.556\,\text{mm}$) and a 5 mm $\times$ 5 mm pad (denoted as Case $5\times5$, measured at $x = 0.460\,\text{mm}$ and $x = 2.545\,\text{mm}$). Error bars on the experimental data indicate measurement uncertainty from the NIST AM-Bench 2025 dataset. The enhanced model incorporates a layer-dependent effective absorptivity ($\eta_{\text{eff}}$) to capture increased laser--material interaction within the powder bed.}
    \label{fig:NIST_result}
\end{figure}

\begin{table}[!h]
\centering
\caption{Summary of simulation and validation cases for laser melting for different cases.}
\label{tab:simulation_cases}
\begin{tabular*}{\textwidth}{@{\extracolsep{\fill}}cccccccc}
\toprule
\textbf{Case} & \textbf{Powder Layer} & \textbf{Pad Size} & \textbf{Position} & \textbf{Spot Diameter} & \textbf{Power, P} & \textbf{Speed, V} & \textbf{VED} \\
 & \textbf{Thickness} ($\mu$m) & (L mm x W mm) & (mm) & ($\mu$m) & (W) & (m/s) & ($J/mm^3$) \\
\midrule
\multicolumn{8}{c}{\textbf{Simulation Cases}} \\
\midrule
1 & 0 & 5$\times$5 & 0.460 & 72 & 285 & 0.96 & 72.9 \\
2 & 0 & 5$\times$5 & 2.545 & 72 & 285 & 0.96 & 72.9 \\
3 & 0 & 1$\times$5 & 0.556 & 72 & 285 & 0.96 & 72.9 \\
\addlinespace 
4 & 80 & 5$\times$5 & 0.460 & 72 & 285 & 0.96 & 72.9 \\
5 & 80 & 5$\times$5 & 2.545 & 72 & 285 & 0.96 & 72.9 \\
6 & 80 & 1$\times$5 & 0.556 & 72 & 285 & 0.96 & 72.9 \\
\addlinespace 
7 & 160 & 5$\times$5 & 0.460 & 72 & 285 & 0.96 & 72.9 \\
8 & 160 & 5$\times$5 & 2.545 & 72 & 285 & 0.96 & 72.9 \\
9 & 160 & 1$\times$5 & 0.556 & 72 & 285 & 0.96 & 72.9 \\
\midrule
\multicolumn{8}{c}{\textbf{Validation Cases \cite{nist_amb2025_results, khorasaniNumericalAnalyticalInvestigation2021}}} \\
\midrule
A & 0 & 0.5$\times$0.5 & 0.25 & 72 & 285 & 0.96 & 72.9 \\
B & 50 & 1$\times$0.5 & 0.25 & 100 & 100 & 1 & 12.7 \\
C & 50 & 1$\times$0.5 & 0.25 & 100 & 370 & 0.8 & 58.9 \\ 
\bottomrule
\end{tabular*}
\parbox{\textwidth}{%
\small 
\vspace{2mm}
Note: 1. For the multi-track simulations: Hatch spacing = 110 $\mu m$ and Laser turn-around time = 0.75 ms is used. 2. Here, L denote length of the simulation domain along the direction of laser scan track and W denote the width of the simulation domain across the direction of laser scan track. The powder layer height for case B and C are extracted from published figures using digital image analysis techniques.
}
\end{table}

\begin{figure}[!h]  
    \centering
    \includegraphics[width=0.8\textwidth]{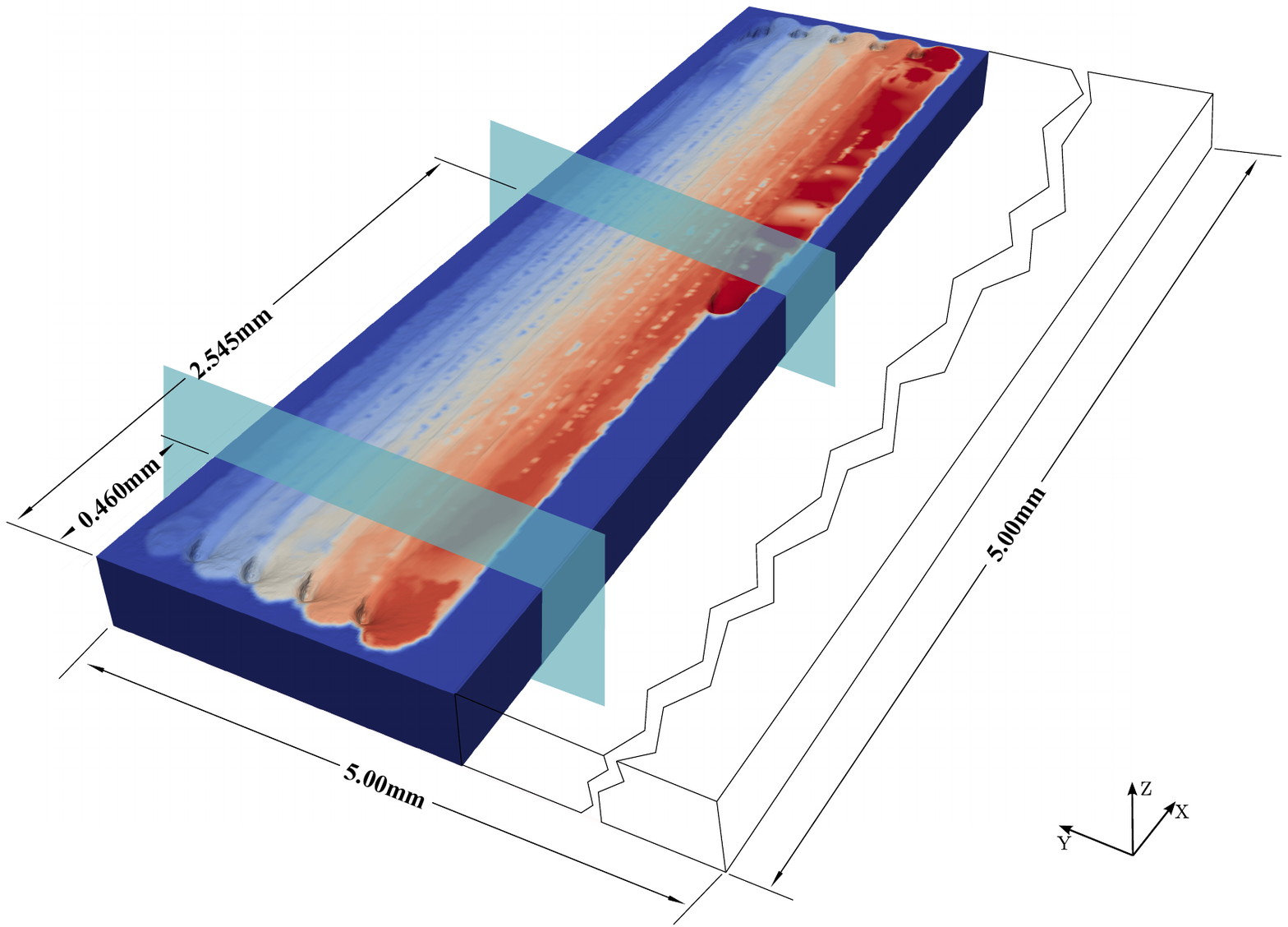}
    \caption{Reference schematic showing the positions of cross-sectional planes in a multi-track scan where measurements are taken. This image corresponds to the 5 mm $\times$ 5 mm pad case, where measurements are performed at two positions ($x = 0.460$ mm and $x = 2.545$ mm). A similar approach is used for the 1 mm $\times$ 5 mm pad case, where the measurement is taken at $x = 0.556$ mm. Different colors represent consecutive tracks.}
    \label{fig:isometric_view}
\end{figure}

Fig. \ref{fig:NIST_result} compares simulated and experimental results at measurement locations specified in Table \ref{tab:simulation_cases} and the laser parameters used in each case specified in Supplementary Table \ref{tab:Laser_settings}. The analysis consists of 9 NIST submission cases with two pad sizes (two different part dimensions) of $1~mm \times 5~mm$ and $5~mm \times 5~mm$. The cases include one bare plate configuration and two powder bed configurations with powder layer heights of 80 and 160 $\mu m$. Measurements were taken at $0.556~mm$ for the $1~mm \times 5~mm$ pad and at $0.460$ and $2.545~mm$ for the $5~mm \times 5~mm$ pad, as shown in Fig. \ref{fig:isometric_view}, across the direction of the laser scan path. For all multitrack cases, the process parameters were power $P = 285~W$, scanning speed $V = 0.96~m/s$, and laser spot diameter of $72~\mu m$. The results compare two distinct modeling iterations against the experimental benchmark. The 'NIST Submission' represents the initial blind predictions performed prior to the release of the NIST dataset. In contrast, the 'Enhanced Model' refers to post-test simulations refined by incorporating additional physical parameters—specifically an effective laser absorptivity scheme—informed by the experimental observations to improve predictive fidelity.

\begin{figure}[!h] 
    \centering
    \includegraphics[width=\textwidth]{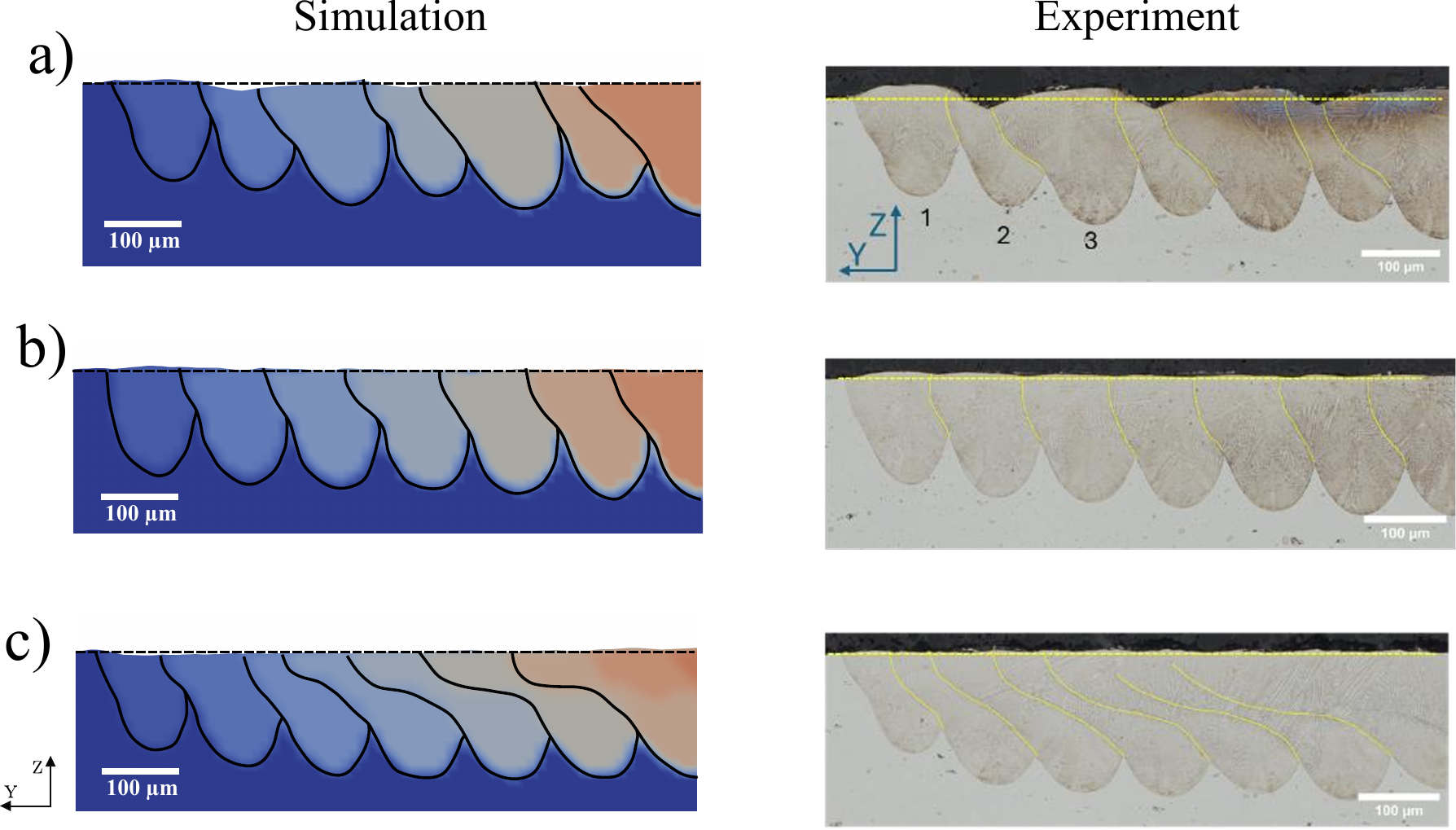}
    \caption{Cross-sections of simulated (left) and experimental (right) melt tracks for bare plates from the AM-Bench 2025 dataset \cite{nist_amb2025_results}. (a) 5 mm $\times$ 5 mm pad at $x = 0.460$ mm, (b) 5 mm $\times$ 5 mm pad at $x = 2.545$ mm, and (c) 1 mm $\times$ 5 mm pad at $x = 0.556$ mm. In the simulation images, distinct colors represent individual scan tracks (Track 1, Track 2, Track 3, etc.), ordered sequentially from left to right. Solid black contour lines delineate the melt-pool boundaries, while the dashed line indicates the initial substrate height.}
    \label{fig:meltpool_bp_image}
\end{figure}

For the bare-plate configurations (Cases 1, 2, and 3), the predicted melt-pool depth, bead height, overlap depth, and dilution layer area demonstrate strong quantitative fidelity, with values remaining within 20\% of experimental measurements as plotted in Fig. \ref{fig:NIST_result}. The results indicate that melt-pool characteristics remain consistent for the 5~mm $\times$ 5~mm pad across different longitudinal positions ($0.460$~mm and $2.545$~mm). Conversely, the smaller pad geometry (1~mm $\times$ 5~mm) exhibits distinct behavior; experimental cross-sections suggest that residual heat from preceding tracks significantly influences melting dynamics, leading to pronounced remelting. As illustrated in the morphological comparison in Fig. \ref{fig:meltpool_bp_image}, the computational model accurately captures these thermal effects and the resulting melt-pool evolution.

\begin{figure}[!h] 
    \centering
    \includegraphics[width=\textwidth]{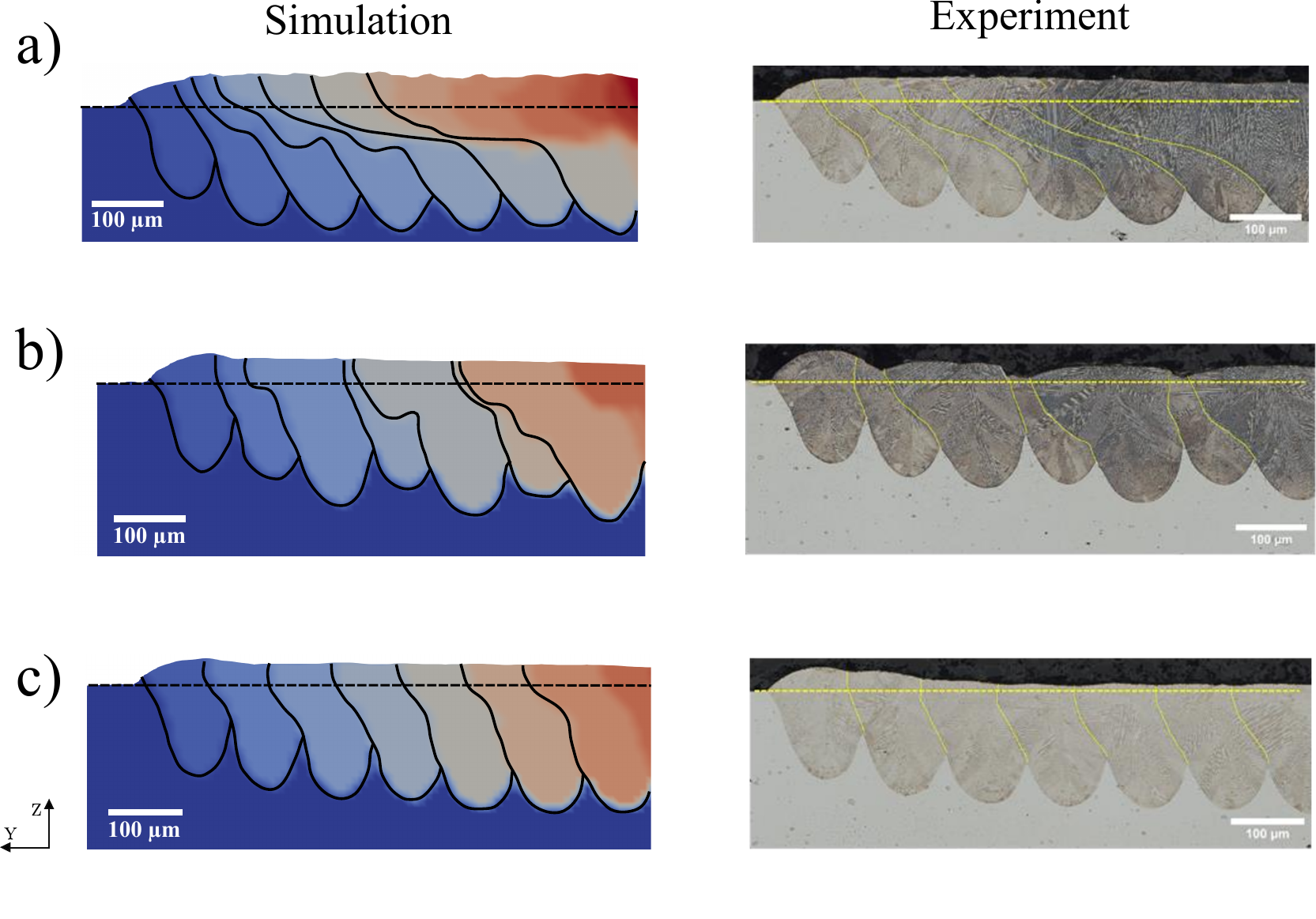}
    \caption{Cross-sections of simulated (left) and experimental (right) melt tracks for an $80\,\mu\text{m}$ powder layer from the AM-Bench 2025 dataset \cite{nist_amb2025_results}. (a) 1 mm $\times$ 5 mm pad at $x = 0.556$ mm, (b) 5 mm $\times$ 5 mm pad at $x = 0.460$ mm, and (c) 5 mm $\times$ 5 mm pad at $x = 2.545$ mm. In the simulation images, distinct colors represent individual scan tracks (Track 1, Track 2, Track 3, etc.), ordered sequentially from left to right. Solid black contour lines delineate the melt-pool boundaries, while the dashed line indicates the initial substrate height.}
    \label{fig:meltpool_80_image}
\end{figure}

\begin{figure}[!h] 
    \centering
    \includegraphics[width=\textwidth]{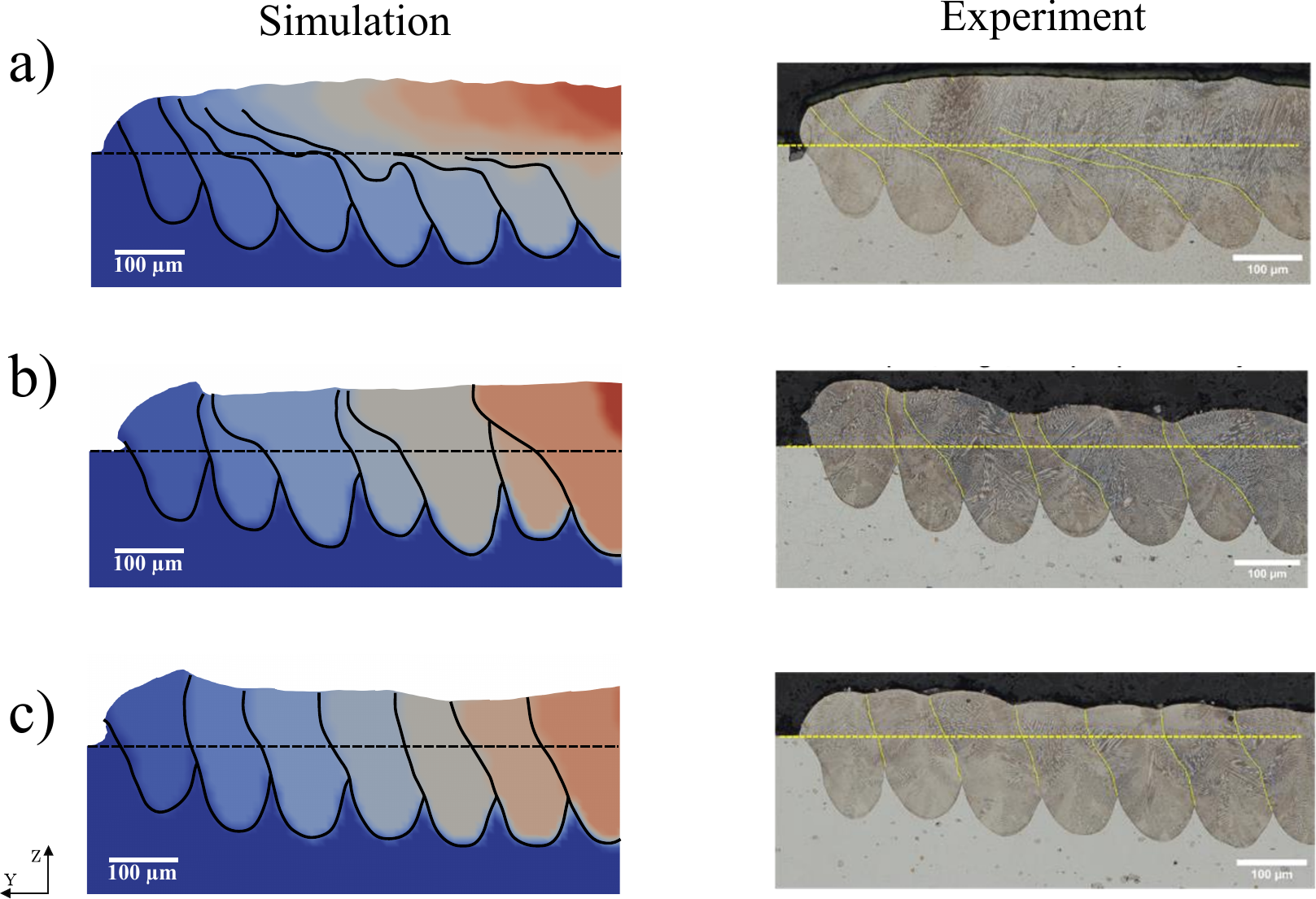}
    \caption{Cross-sections of simulated (left) and experimental (right) melt tracks for a $160\,\mu\text{m}$ powder layer from the AM-Bench 2025 dataset \cite{nist_amb2025_results}. (a) 1 mm $\times$ 5 mm pad at $x = 0.556$ mm, (b) 5 mm $\times$ 5 mm pad at $x = 0.460$ mm, and (c) 5 mm $\times$ 5 mm pad at $x = 2.545$ mm. In the simulation images, distinct colors represent individual scan tracks (Track 1, Track 2, Track 3, etc.), ordered sequentially from left to right. Solid black contour lines delineate the melt-pool boundaries, while the dashed line indicates the initial substrate height.}
    \label{fig:meltpool_160_image}
\end{figure}


As illustrated in Fig. \ref{fig:NIST_result}, the NIST submission model exhibited limited predictive accuracy for melt-pool dimensions in both the 80~$\mu$m and 160~$\mu$m powder-bed configurations. This discrepancy underscores the importance of incorporating a layer-dependent effective absorptivity model, which was implemented in the enhanced model to more accurately capture the laser--material interaction within the powder bed.

During the 'PowderSim' variable turned on in the solver, the solver is modified to prevent individual, partially melted powder particles from behaving like liquid. This is achieved by switching from the standard liquid fraction to a masked liquid fraction for calculating the Darcy momentum sink. The mask treats a cell as liquid only if its local average liquid fraction exceeds a threshold of 0.95. This approach prevents flow of partially melted particles and ensures that flow begins only once a contiguous melt pool forms. Additionally, gravitational forces are selectively applied only to the fully formed liquid melt pool while ignoring individual solid powder particles. With this approach, the laser absorptivity calibration parameters (notably the electric resistivity and electron number density) were kept unchanged. Maintaining these settings constant across both bare plate and powder bed cases was hypothesized to result in accurate predictions for powder bed configurations. However, these physics modifications resulted in reduced prediction accuracy of melt pool dimensions in the NIST submission model. 

This approach necessitated the development of an enhanced model that can predict accurately across both powder bed and bare-plate cases consistently. Since the bare plate cases already yielded highly accurate results and the modifications in the enhanced model do not affect bare plate simulations, only powder bed cases are simulated with the `Enhanced Model'. 

In this enhanced model, the \texttt{powderSim} variable was set to false, matching the bare plate configuration. This disables the default porosity models that automatically adjust thermal conductivity, bulk density, and electrical resistivity. Instead, heat absorptivity is manually controlled, and distinct material properties (bulk density, thermal conductivity, specific heat, and electrical resistivity) are assigned directly to the powder domain. Since powder particles exhibit lower thermal conductivity and higher laser reflection compared to the solid substrate \cite{quevaNumericalStudyImpact2020}, they effectively absorb more energy; therefore, the absorptivity was formulated as a function of the powder layer thickness. In the interest of simulation convergence, the powder material was initially treated with properties similar to the solid substrate. However, as the thermophysical properties differ between powder particles and bulk material, these effects were incorporated through adjustments in electrical resistivity ($R_e$), which modulates the laser energy absorption accordingly. To implement the calculated layer-dependent effective absorptivity ($\eta_{\text{eff}}$) within the ray-tracing framework, $R_e$ was utilized as the primary calibration parameter. In the Drude-model-based Fresnel formulation used by the solver (Eqs.~\ref{eq:er}--\ref{eq:ei}), $R_e$ directly modulates the damping frequency ($\delta$), which in turn dictates the complex relative permittivity and the resulting energy absorption per ray interaction. The specific $R_e$ values listed in Table~\ref{tab:Laser_settings} were determined through an iterative inverse-calibration process: for each powder layer height, $R_e$ was adjusted until the predicted melt-pool dimensions---specifically depth, width, and bead height---demonstrated the best agreement with the NIST experimental cross-sections. This methodology allows the model to macroscopically account for the increased energy coupling caused by multiple reflections within the porous powder bed and the differences in thermophysical coupling between the bulk material and discrete particles.

In order to develop a scheme for average laser absorption for powder layer, we started studying the laser absorption for different scenarios (bare plate and powder layer of different heights). For the bareplate validation case A, the average absorptivity was found to be 0.51 (Appendix Fig. \ref{fig:validation_abs}). Since the process parameters remain unchanged across all cases (bareplate, 80~$\mu m$, and 160~$\mu m$ powder bed), the absorptivity can be treated primarily as a function of layer thickness $L$. The relationship is not linear; rather, it follows an asymptotic trend as the bed transitions from a `transparent' thin layer for bare plate, where the substrate properties dominate, to a `semi-infinite' thick layer for powder layers, where the bed properties saturate~\cite{hondaMeasurementLaserAbsorptivity2025,simondsDynamicLaserAbsorptance2020}. We propose, this relationship can be approximated by an exponential saturation function of the form:

\begin{equation}
    \eta_{\text{eff}}(L) = \eta_{\text{solid}} + (\eta_{\infty} - \eta_{\text{solid}}) \cdot (1 - \exp(-\beta L)),
\end{equation}

\noindent where $\eta_{\text{eff}}$ is the effective absorptivity for a given layer thickness $L$, $\eta_{\text{solid}}$ is the base-level absorptivity for $L = 0$ (bare plate), $\eta_{\infty}$ is the asymptotic limit where the absorptivity value saturates, and $\beta$ is the decay constant.

For the bare plate case, the average absorptivity is taken as $0.51$. For IN718, a maximum absorptivity of $0.87$ has been reported for the LPBF process~\cite{romanoLaserAdditiveMelting2016}, so $\eta_{\infty}$ is set to $0.87$. The only remaining unknown is the $\beta$ coefficient. For validation Case C, the volumetric energy density (VED = $58.9$~J/mm$^3$) is close to that of simulation Cases 1--9 ($72.9$~J/mm$^3$), and the mean absorptivity is $0.7836$ (Appendix Fig.~\ref{fig:validation_abs}). Given that the VED of simulation Cases 1--9 ($72.9$~J/mm$^3$) is approximately 24\% higher than that of validation Case C ($58.9$~J/mm$^3$), and absorptivity is known to increase with VED~\cite{ganUniversalScalingLaws2021,aggarwalInvestigationTransientCoupling2023}, the absorptivity for $L = 50$~$\mu$m is conservatively estimated as $0.785$ under the present process conditions, representing a modest upward adjustment from the measured mean of $0.7836$. It is acknowledged that this estimate introduces a small uncertainty in the predicted absorptivity values for $80$~$\mu$m and $160$~$\mu$m layers. Using this value, $\beta$ is determined to be approximately $0.02887$~$\mu$m$^{-1}$, giving:

\begin{equation}
    \eta_{\text{eff}}(L) = 0.51 + (0.87 - 0.51) \cdot (1 - \exp(-0.02887 L)).
\end{equation}

Using this equation, the approximated absorptivity values for $80$~$\mu$m and $160$~$\mu$m are calculated to be $0.834$ and $0.866$, respectively. 

\begin{figure}[!h]
    \centering
    \includegraphics[width=0.95\linewidth]{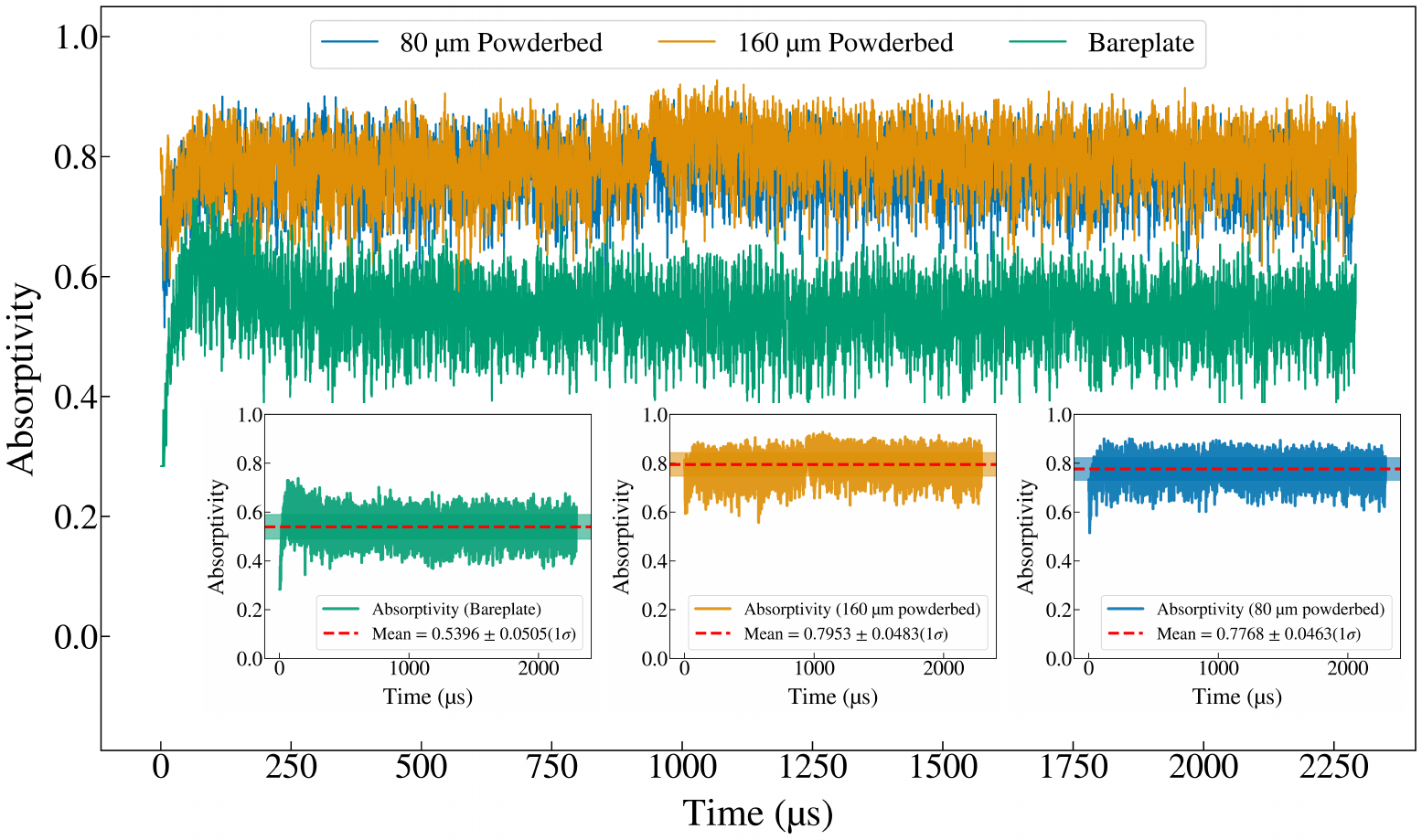}
    \caption{Temporal evolution of laser absorptivity for single-track laser powder bed fusion under three surface conditions: 80~$\mu$m powderbed, 160~$\mu$m powderbed, and bareplate. The main plot shows all three cases overlaid for direct comparison, while the insets present individual traces with the mean absorptivity (red dashed line) and $\pm1\sigma$ band. Process parameters were constant for all cases: laser power $P = 285$~W, scanning speed $v = 0.96$~m/s, spot diameter $d = 72$~$\mu$m, and track length $L = 2.5$~mm.}
    \label{fig:absorptivity_comparison}
\end{figure}

To verify that these analytical approximations align with the physical behavior of the melt pool, high-fidelity single-track simulations ($2.5$~mm track length) were conducted for the bare plate, $80$~$\mu$m, and $160$~$\mu$m configurations using the same laser parameters specified in Table~\ref{tab:Laser_settings}. As shown in Figure~\ref{fig:absorptivity_comparison}, the time-averaged absorptivity measured directly from the simulations was $0.540 \pm 0.050$ for the bare plate, $0.777 \pm 0.046$ for the $80$~$\mu$m powder bed, and $0.795 \pm 0.048$ for the $160$~$\mu$m powder bed. These simulation results strongly support the proposed saturation scheme. The bare plate exhibits the lowest absorption because a significant portion of the incident rays are reflected directly from the flat surface; in contrast, the $80$~$\mu$m powder layer significantly enhances energy coupling through multiple internal reflections among the powder particles. Notably, further increasing the layer thickness to $160$~$\mu$m yields only a marginal increase in absorptivity, demonstrating the predicted asymptotic behavior. This confirms that at higher thicknesses, the powder bed transitions into a semi-infinite regime where the effective absorptivity saturates. Finally, it is observed that the analytical equation consistently predicts slightly higher absorptivity values than those obtained through high-fidelity simulations ($0.834$ vs. $0.777$ for $80$~$\mu$m, and $0.866$ vs. $0.795$ for $160$~$\mu$m). Consequently, the analytical relationship serves as a conservative upper bound for energy coupling in the multi-track simulation framework.

The enhanced model demonstrates noticeable improvement in predicting melt pool depth, width, bead height, overlap depth, and solidified layer areas for powder layer height of 80 $\mu$m, shown in Fig. \ref{fig:meltpool_80_image} and powder layer height of 160 $\mu$m in Fig. \ref{fig:meltpool_160_image}, where each track is visualized in distinct colors. Similar to the bareplate case of Fig. \ref{fig:meltpool_bp_image}, we observe the effect of residual heat and remelting for pad size of $1 \times 5 mm$ for the $80 \mu m$ cases. The effect of residual heat influencing the melt pool dynamics is also observed for $5 \times 5 mm$ at two different location. The cross-section image at 0.460 $mm$ shows the dominant melt pool at alternate odd tracks. We hypothesize that the elevated temperature of the even tracks influenced the remelting of the next odd track as it is less time to cool off because of the cross-section location being closer to the starting iposition. On the other hand, when the cross-section is taken at 2.545 $mm$, the tracks had sufficient time to stabilize which resulted in a more symmetric appearance of melt pool morphology (Fig. \ref{fig:meltpool_80_image}). This trend continues for $160 \mu m$ powder layer height and the experimental observation is picked up in the computational model as well. As experimental microgrpahs were only shared for first 7 tracks by NIST, we also carried out our simulation up to 15 tracks. Also, due to computational limitations, approximately it is significantly challenging to simulate all 45 tracks. Fig. \ref{fig:5x5_bp_2.545}, \ref{fig:5x5_80_2.545}, \ref{fig:5x5_160_2.545} shows the variation in melt pool dimensions with increasing track numbers for multi-track laser scan at the mid-location of the 5$\times$5 mm pad for three powder layer thicknesses.

We also observe that if the process parameters are kept constant i.e. no change in laser power, scan speed, spot diameter, hatch spacing, turnaround time, and layer height; the melt pool morphology stabilizes and demonstrate a steady change in the subsequent laser tracks after track 10. Picking up this trend as all the process variations were kept unchanged, an extrapolation method was used to predict data up to the 45th track. An exponential fit was used for depth, width, bead height, and overlap depth, as this functional form best supports the trends observed in AM-Bench 2022 results \cite{amb2022results}. For dilution layer area and solidified layer area, a linear fit was applied since these represent cumulative quantities that increase progressively with each track. The results are presented in \ref{appendix:results} for comparison. 

\begin{figure}[!h]
    \centering
    \includegraphics[width=\textwidth]{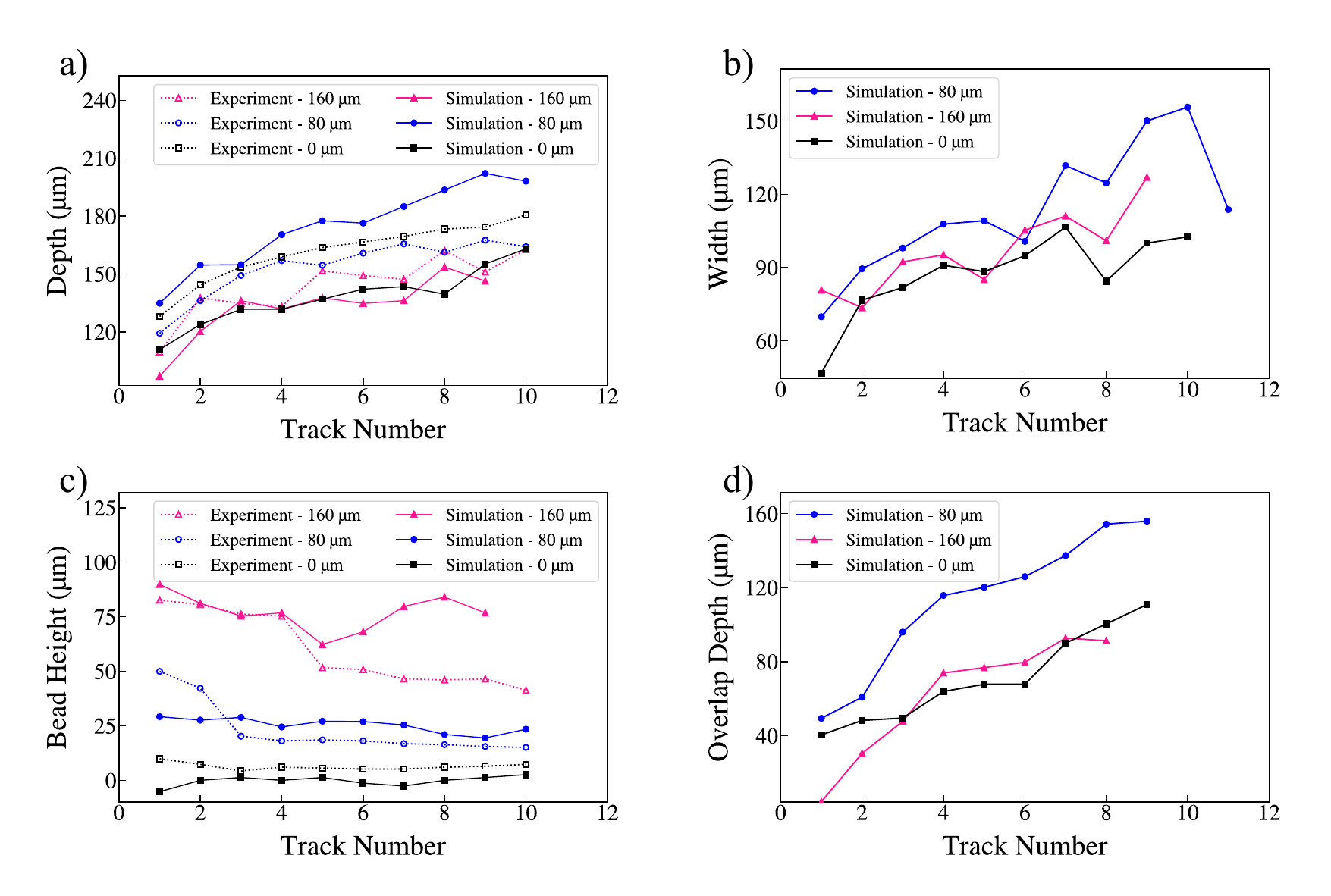}
    \caption{Variation of melt-pool geometric characteristics across sequential laser scan tracks for the $5 \times 5$~mm pad geometry at the longitudinal mid-section ($x = 2.545$~mm). Panels illustrate (a) melt-pool depth, (b) melt-pool width, (c) bead height, and (d) overlap depth for three powder layer configurations (0~$\mu$m (bareplate), 80~$\mu$m, and 160~$\mu$m). Note that experimental benchmark data for track-by-track comparisons were unavailable for the metrics presented in (b) and (d).}
    \label{fig:variation_metrics}
\end{figure}

Although the extrapolated predictions using the enhanced model were in closer agreement with experimental results (within 20\% of the experiment for depth and dilution area), minor discrepancies remain, particularly in bead height and solidified layer area. This difference may be attributed to the fact that the residual heat factor on the laser scan track remains challenging to fully resolve within the current modeling framework without high-fidelity simulation. The actual simulation of all the 45 track is expected to provide improved accuracy of the temperature evolution and will thus result in more accurate melt-pool morphologies. Considering this effect of temperature more accurately would result in deeper melt pool penetration and consequently lower bead height, which would better match the experimental observations. Next, the over-prediction of bead height directly contributes to the higher solidified layer area estimation in the simulations when compared to experiment. 

Interestingly, the overlap depth shows some disagreement with experimental data in $5 \times 5 mm$ powder-beds at measurement position $x = 0.46 mm$ layer thickness where in other layer thickness cases the error margin falls mostly under 15\%. This behavior is associated with the current extrapolation approach, which is based on data from only 10 simulated tracks.

Given the high computational cost of high-resolution simulations, a mesh size of 10 $\mu$m was employed. Finer mesh resolution may improve the accuracy of the predictions of overlap depth and width by better resolving the steep temperature and velocity gradients in the melt pool region. However, refining the mesh significantly increases the computational cost and simulation time, which limits the feasibility of simulating the full 45-track model at finer resolutions.

In summary, the enhanced model shows substantial improvement over the NIST submission model for different powder layer heights and of different part length while maintaining good accuracy. Further improvements in predictive accuracy can be achieved through increased computational resources or the integration of reduced-order or data-driven modeling approaches.

\section*{Conclusions}

This study aimed to develop a high-fidelity multiphysics simulation framework using the OpenFOAM based solver `LaserBeamFoam' to investigate the melt pool dynamics of IN718 during the Laser Powder Bed Fusion (LPBF) process across varying powder layer thickness. The findings of the study show that the simulated melt pool dimensions achieved excellent quantitative agreement with the highly controlled NIST AM-Bench 2025 experimental data for both bare plate and powder bed configurations. Furthermore, by systematically adjusting the effective laser absorptivity as a function of the powder layer thickness, the model successfully extrapolated multi-track geometrical characteristics up to the 45th consecutive laser track. The novelty of these findings lies in the systematic application of this layer-dependent effective absorptivity approach within a unified simulation environment to consistently predict extended multi-track geometries. To the best of our knowledge, this is the first ever computational approach to predict melt pool morphologies for varying powder layer thicknesses and different part dimensions which agrees reasonably well with the rigorous AM-Bench 2025 experiments.
An interesting approach of the current study is to account for the changing powder layer properties through the use of changing laser absorption. We introduced an effective average laser absorption scheme for multi-track simulation to accelerate the computational method. With this approach, this study makes an important contribution to establish highly accurate, foundational methodology for multi-track predictions that can be seamlessly utilized in future studies through the implementation of finer mesh resolutions. Ultimately, it establishes a verified pathway toward the integration of predictive multi-physics simulations into advanced digital twin frameworks for additive manufacturing.

\section*{Author contributions}
\textbf{Badhon Kumar:} Conceptualization, Methodology, Formal analysis, 
Writing – original draft, Writing – review and editing. 
\textbf{Rakibul Islam Kanak:} Writing – review and editing. 
\textbf{Nishat Sultana:} Writing – original draft, Writing – review and editing. 
\textbf{Jiachen Guo:} Writing – review and editing. 
\textbf{Andrew Schrader:} Writing – review and editing. 
\textbf{Wing Kam Liu:} Writing – review and editing. 
\textbf{Abdullah Al Amin:} Conceptualization, Resources, 
Writing – original draft, Writing – review and editing.

\section*{Conflicts of interest}
The authors declare no conflicts of interest.

\section*{Data availability}
Data is available upon reasonable request.

\section*{Acknowledgements}

The authors acknowledge gracious support from the Ohio Supercomputer Center through academic project and Anvil Supercomputer Center at Purdue University through NSF ACCESS for the high performance computation resources. Abdullah Al Amin acknowledges support from the School of Engineering at the University of Dayton through the Research Council Seed Grant and National Science Foundation Award \# 2501711.

\section*{Declaration of Generative AI and AI-assisted technologies in the writing process}
During the preparation of this work the authors used ChatGPT, Claude, and 
Gemini in order to improve readability and language. After using these tools, 
the authors reviewed and edited the content as needed and take full 
responsibility for the content of this publication.





\appendix
\renewcommand{\thefigure}{\Alph{section}\arabic{figure}}
\renewcommand{\thetable}{\Alph{section}\arabic{table}}
\renewcommand{\theequation}{\Alph{section}\arabic{equation}}
\section{Supplementary Data}
\label{appendix:methods}
\setcounter{figure}{0}
\setcounter{table}{0}
\setcounter{equation}{0}

The thermophysical properties of IN718 and Argon gas are presented in Table~\ref{tab:properties}, and the laser parameters employed in the solver across all simulation cases are summarised in Table~\ref{tab:Laser_settings}.

\begin{table}[htbp]
\centering
\caption{Thermophysical properties of Inconel 718 (IN718) and argon gas used in the numerical simulations.}
\label{tab:properties}
\begin{tabular}{llcc}
\hline
\textbf{Property} & \textbf{Symbol / Unit} & \textbf{IN718} & \textbf{Argon} \\
\hline
Density                        & $\rho$ (kg\,m$^{-3}$)           & 7600                & 1.622               \\
Kinematic viscosity            & $\nu$ (m$^2$\,s$^{-1}$)         & $6.7\times10^{-7}$  & $1.53\times10^{-5}$ \\

Surface tension                & $\sigma$ (N\,m$^{-1}$)          & 1.8                 & --                  \\
Surface tension gradient       & $\partial\sigma/\partial T$ (N\,m$^{-1}$\,K$^{-1}$) & $-3.7\times10^{-4}$ & -- \\
Thermal expansion coefficient  & $\beta$ (K$^{-1}$)              & $1.3\times10^{-5}$  & $4.0\times10^{-5}$  \\
Solidus temperature            & $T_s$ (K)                        & 1533                & --                  \\
Liquidus temperature           & $T_l$ (K)                        & 1609                & --                  \\
Latent heat of fusion          & $L_f$ (J\,kg$^{-1}$)            & $2.50\times10^{5}$  & --                  \\
Vaporisation temperature       & $T_\text{vap}$ (K)              & 3000                & --                  \\
Latent heat of vaporisation    & $L_v$ (J\,kg$^{-1}$)            & $7.34\times10^{6}$  & --                  \\
Molar mass                     & $M_m$ (kg\,mol$^{-1}$)          & 0.0585              & 0.03995             \\
Ambient pressure               & $p_0$ (Pa)                       & \multicolumn{2}{c}{101325}              \\
Thermal conductivity           & $\kappa$ (W\,m$^{-1}$\,K$^{-1}$) & see Fig.~\ref{fig:tabular-thermal-prop} & $1.7\times10^{-2}$ \\
Specific heat capacity         & $c_p$ (J\,kg$^{-1}$\,K$^{-1}$)   & see Fig.~\ref{fig:tabular-thermal-prop} & 520                \\
\hline
\end{tabular}
\end{table}

\begin{figure}[h!]
    \centering
    \includegraphics[width=\textwidth]{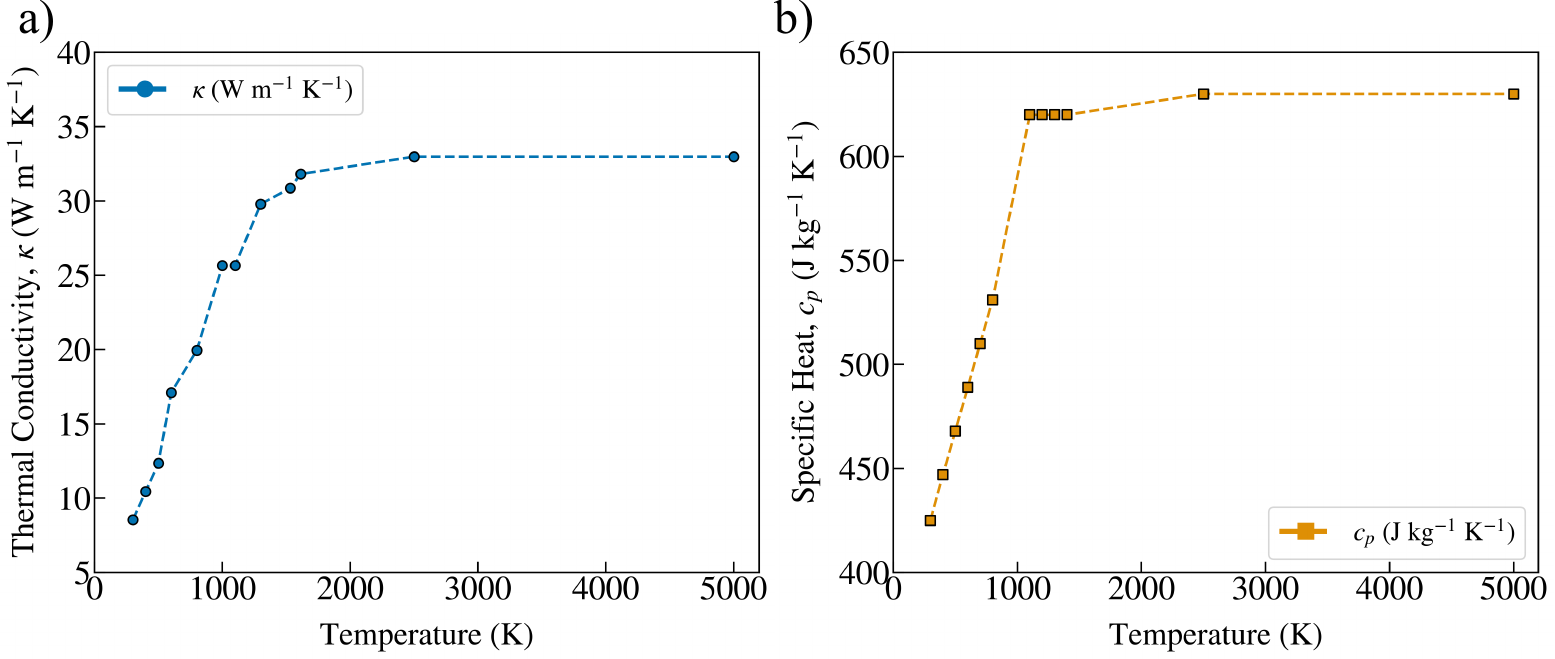}
    \caption{Thermophysical properties of Inconel 718 (IN718) as a function of temperature:
    (a) thermal conductivity $\kappa$ (W\,m$^{-1}$\,K$^{-1}$) and
    (b) specific heat capacity $c_p$ (J\,kg$^{-1}$\,K$^{-1}$).}
    \label{fig:tabular-thermal-prop}
\end{figure}

\begin{figure}[!h]  
    \centering
    \includegraphics[width=\textwidth]{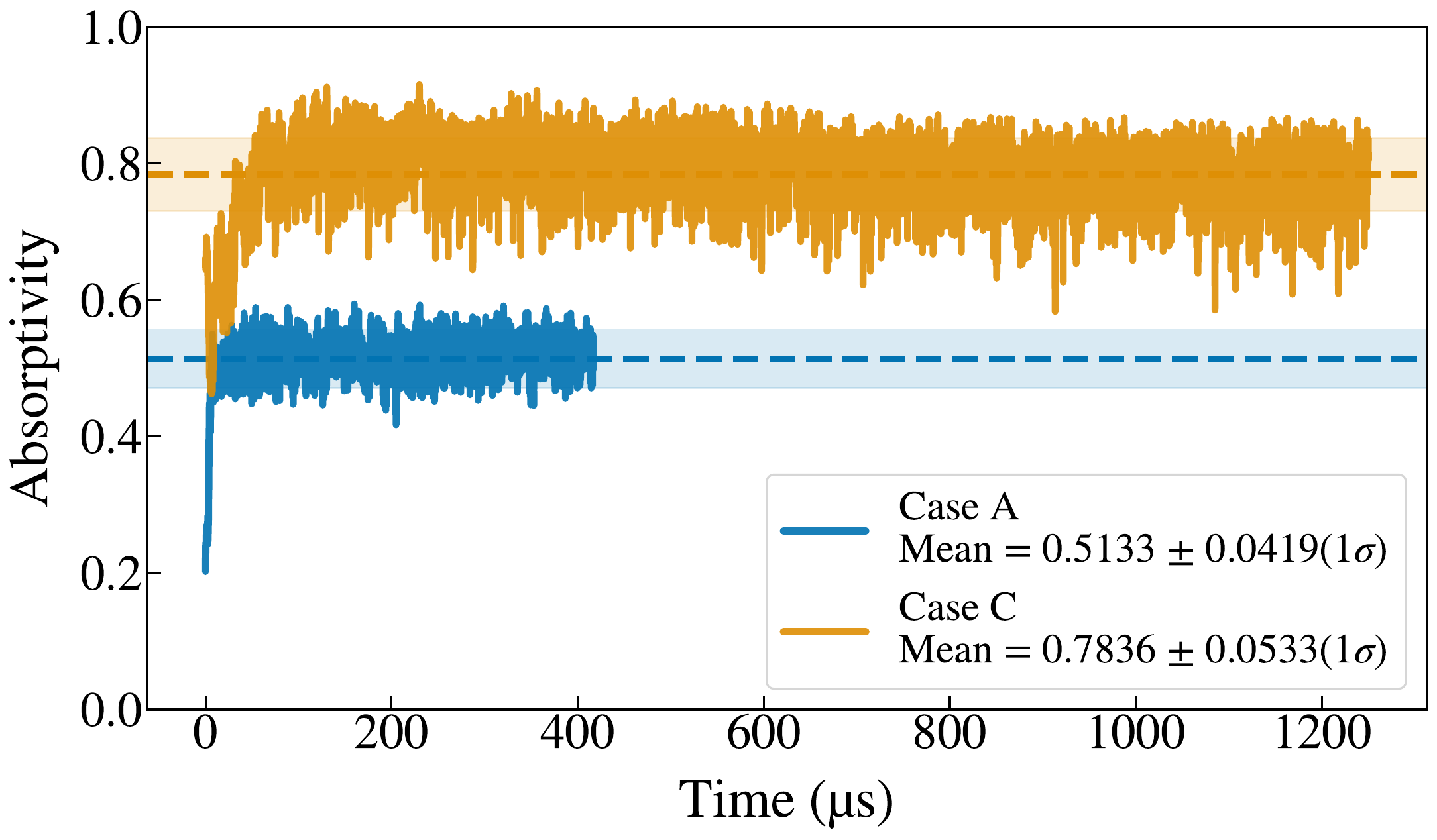}
    \caption{Absorptivity as a function of time for validation Cases A and C. Case A corresponds to a bare plate ($0\,\mu$m powder layer, $P = 285\,$W, $V = 0.96\,$m/s, $d = 72\,\mu$m) and Case C to a $50\,\mu$m powder layer ($P = 370\,$W, $V = 0.8\,$m/s, $d = 100\,\mu$m). Dashed lines indicate the temporal mean, and shaded bands denote the $\pm1\sigma$ interval.}
    \label{fig:validation_abs}
\end{figure}

\begin{table}[!h]
    \centering
    \small
    \caption{Laser parameters and material-specific properties used in the simulations.}
    \label{tab:Laser_settings}
    \begin{tabular*}{\textwidth}{@{\extracolsep{\fill}} l c c c c c c}
        \toprule
        & & & \multicolumn{4}{c}{\textbf{Layer thickness}} \\
        \cmidrule{4-7}
        \textbf{Property} & \textbf{Symbol} & \textbf{Unit} & \textbf{Bare plate} & \textbf{50 \textmu m} & \textbf{80 \textmu m} & \textbf{160 \textmu m} \\
        \midrule
        Spot diameter           & $d$        & \si{\micro\meter} & 72 & 100 & 72 & 72 \\
        Wavelength              & $\lambda$  & \si{\micro\meter} & 1.064 & 1.064 & 1.064 & 1.064\\
        Electric resistivity    & $R_e$   & \si{\ohm\cdot m}  & $1.1 \times 10^{-6}$ & $5 \times 10^{-6}$ & $5.78 \times 10^{-6}$ & $7.7 \times 10^{-6}$\\ 
        Electron number density & $N_e$       & \si{m^{-3}}       & $1.8 \times 10^{29}$ & $5 \times 10^{29}$ & $5 \times 10^{29}$ & $5 \times 10^{29}$\\
        \bottomrule
    \end{tabular*}
\end{table}
Figure~\ref{fig:validation_abs} shows the simulated absorptivity as a function of time for the validation cases (Cases A and C), for which the melt pool dimensions were validated against experimental measurements.


\clearpage
\section{Supplementary Results}
\label{appendix:results}
\setcounter{figure}{0}
\setcounter{table}{0}
\setcounter{equation}{0}

\begin{figure}[!h]  
    \centering
    \includegraphics[width=\textwidth]{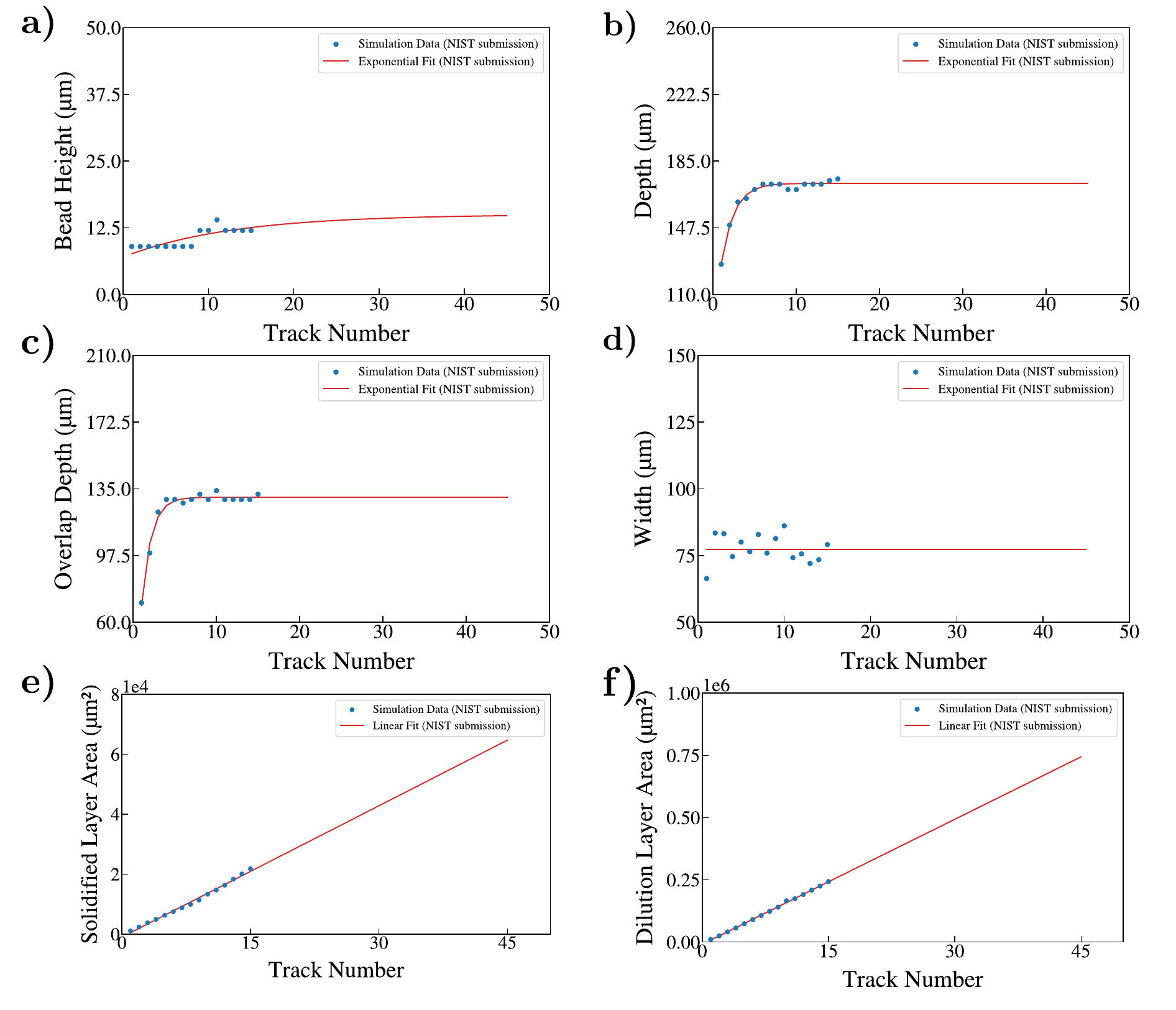}
    \caption{Bare plate prediction (1x5 mm) of 45 track average data using exponential fit measured at x=0.556 mm}
    \label{fig:1x5_bp}
\end{figure}

\begin{figure}[!h]  
    \centering
    \includegraphics[width=\textwidth]{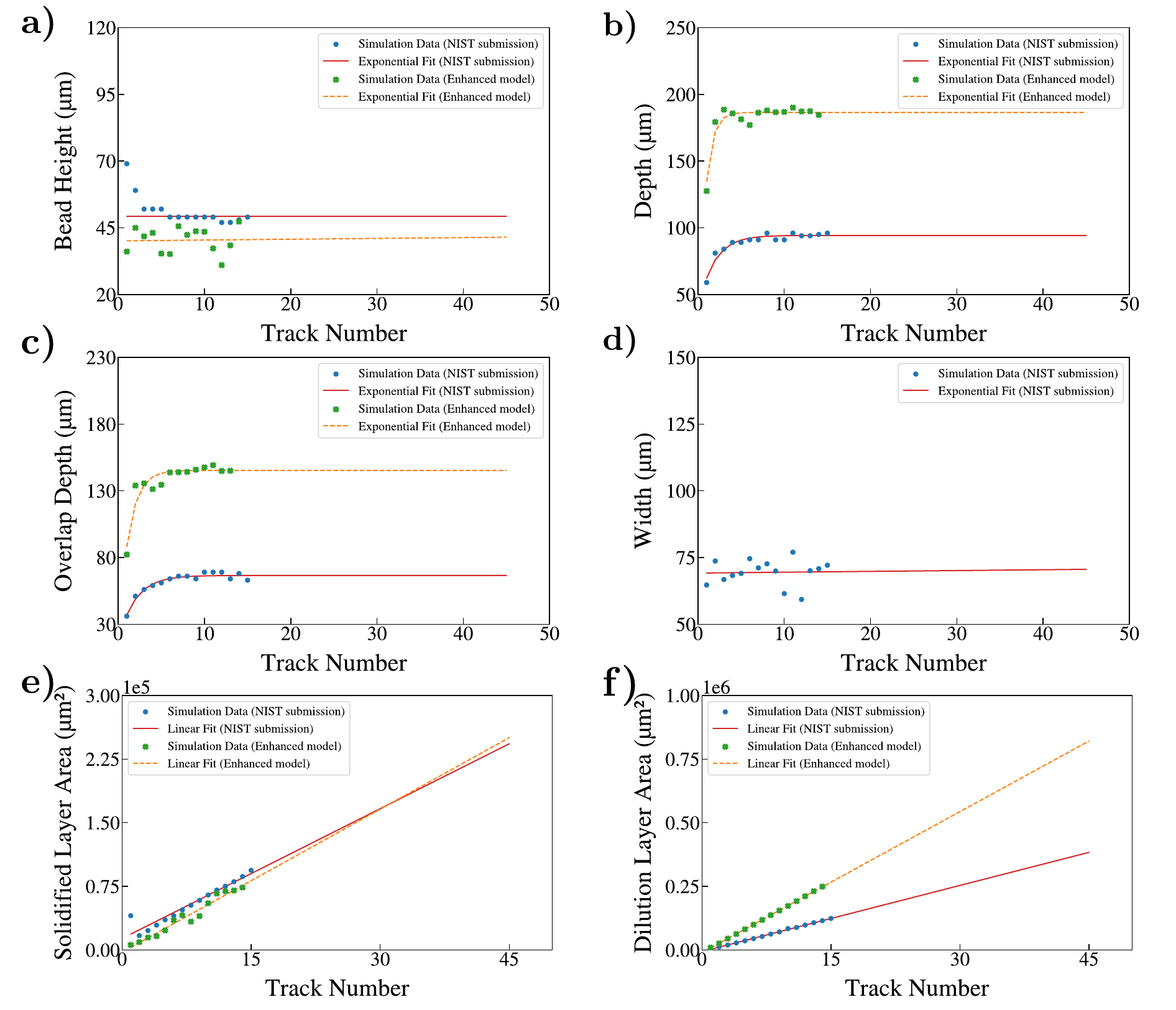}
    \caption{Powder bed of 80 \(\mu\)m prediction (1x5 mm) of 45 track average data using exponential fit measured at x=0.556 mm}
    \label{fig:1x5_80}
\end{figure}

\begin{figure}[!h]  
    \centering
    \includegraphics[width=\textwidth]{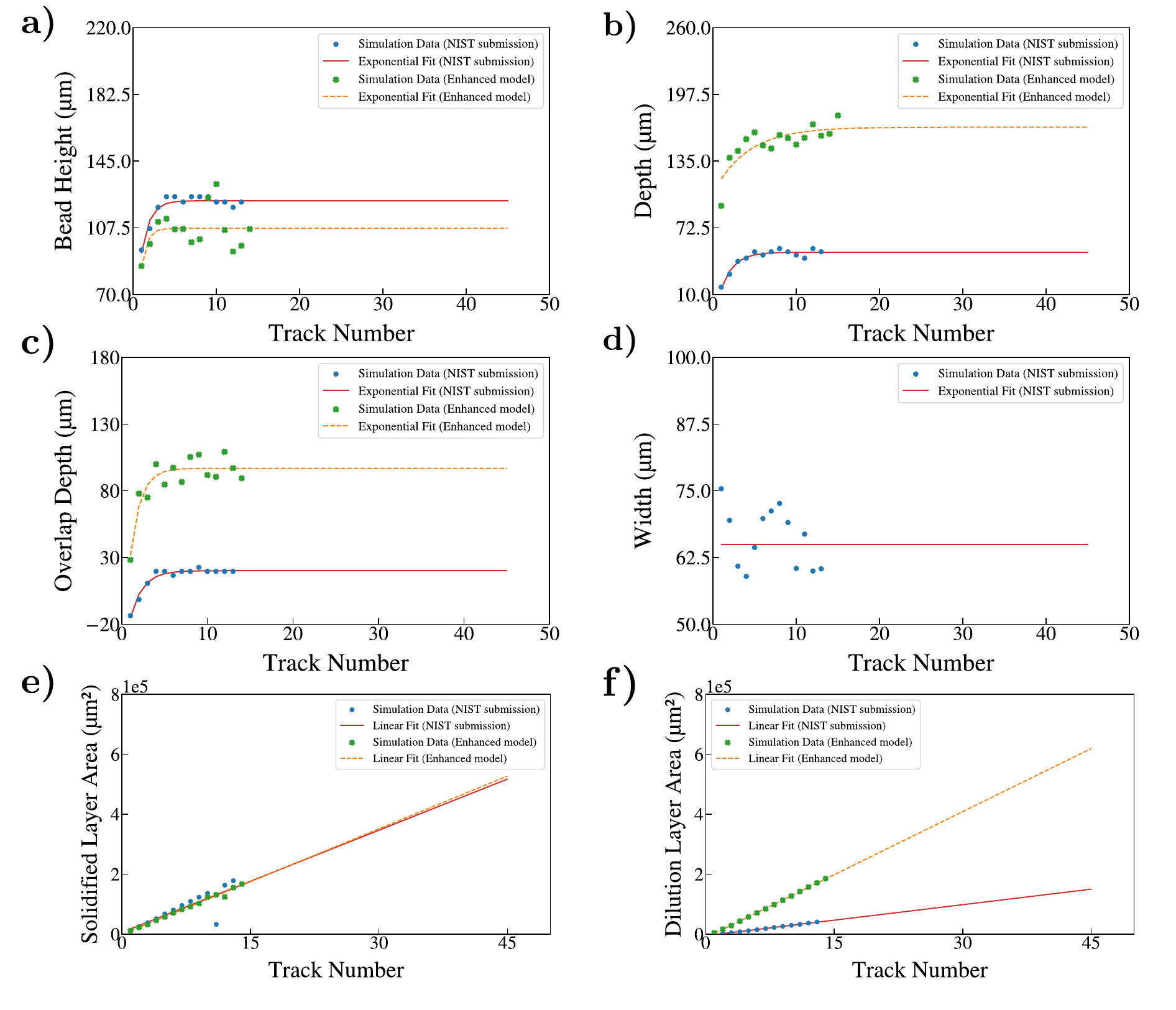}
    \caption{Powder bed of 160 \(\mu\)m prediction (1x5 mm) of 45 track average data using exponential fit for measured at x=0.556 mm}
    \label{fig:1x5_160}
\end{figure}

\begin{figure}[!h]  
    \centering
    \includegraphics[width=\textwidth]{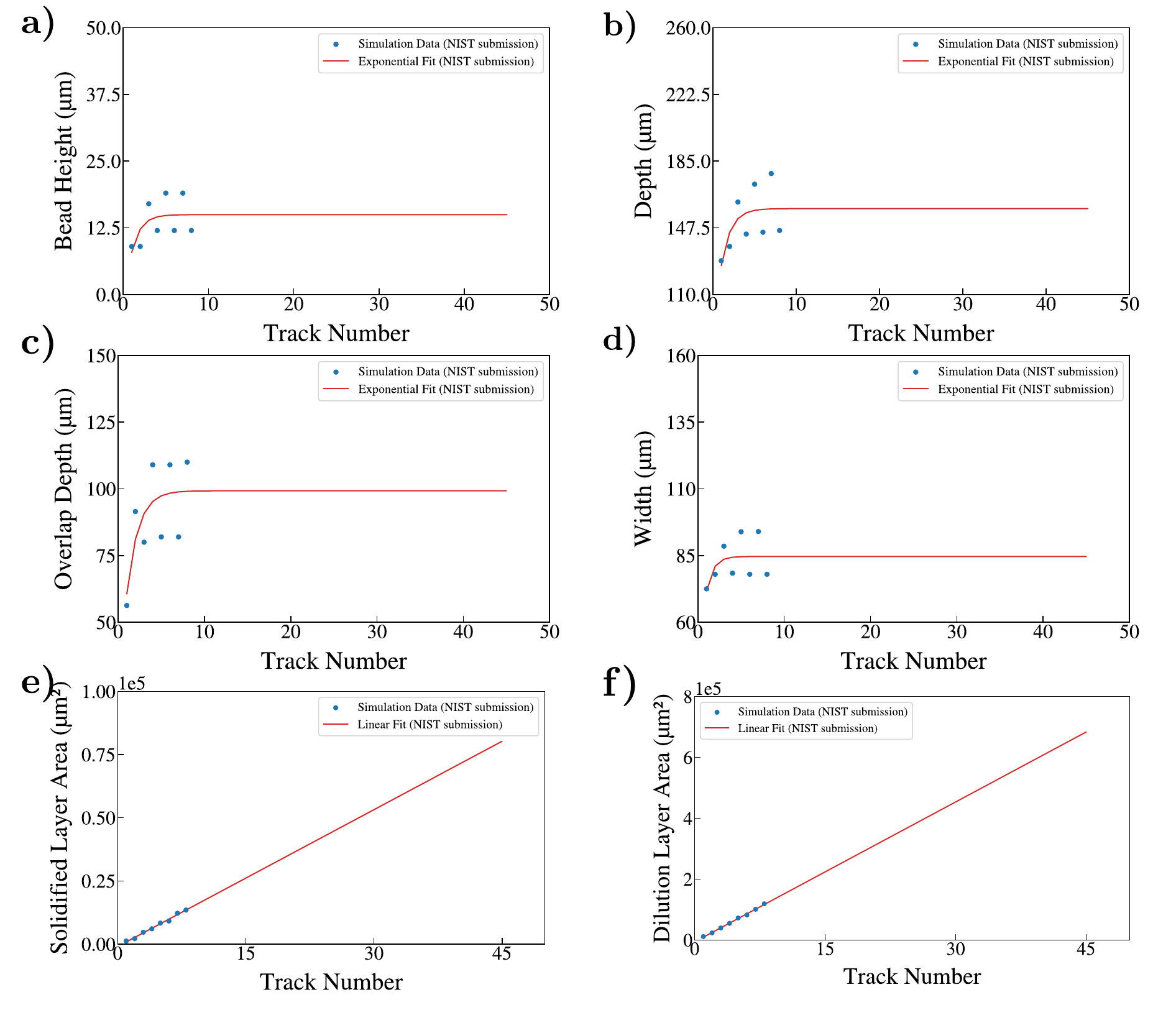}
    \caption{Bare plate prediction (5x5 mm) of 45 track average data using exponential fit measured at x=0.460 mm}
    \label{fig:5x5_bp_0.46}
\end{figure}

\begin{figure}[!h]  
    \centering
    \includegraphics[width=\textwidth]{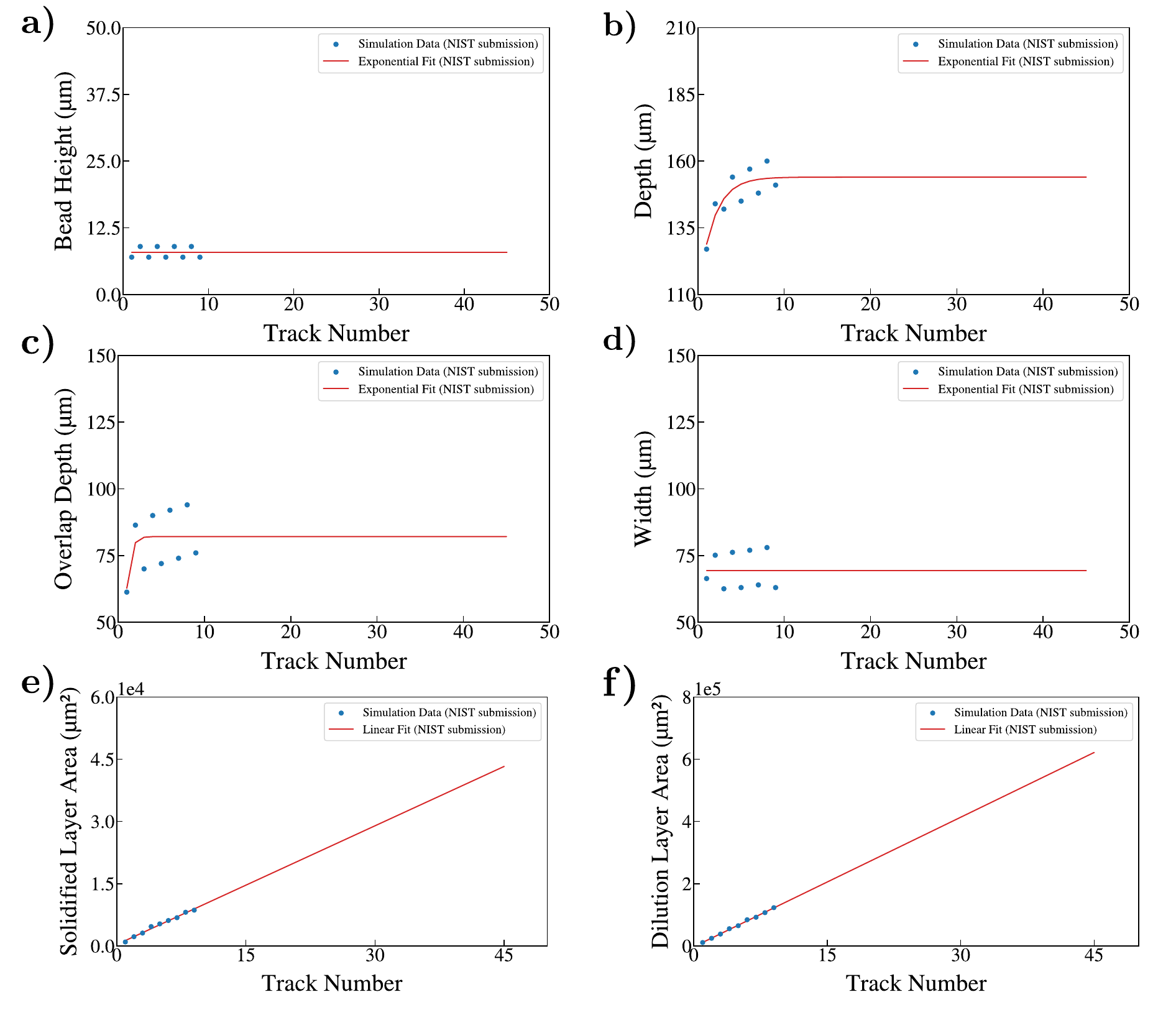}
    \caption{Bare plate prediction (5x5 mm) of 45 track average data using exponential fit measured at x=2.545 mm}
    \label{fig:5x5_bp_2.545}
\end{figure}

\begin{figure}[!h]  
    \centering
    \includegraphics[width=\textwidth]{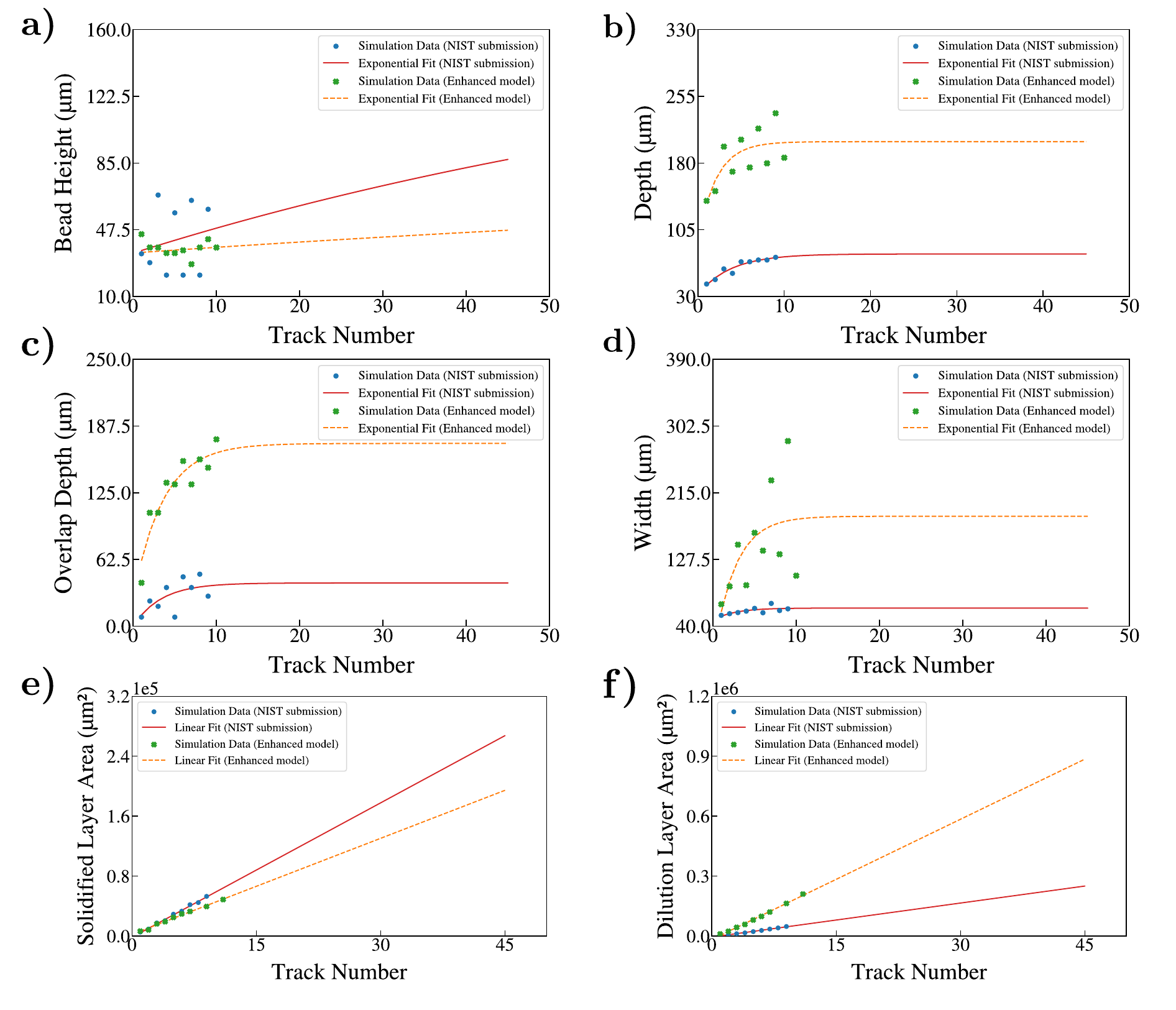}
    \caption{Powder bed of 80 \(\mu\)m prediction (5x5 mm) of 45 track average data using exponential fit for measured at x=0.460 mm}
    \label{fig:5x5_80_0.46}
\end{figure}

\begin{figure}[!h]  
    \centering
    \includegraphics[width=\textwidth]{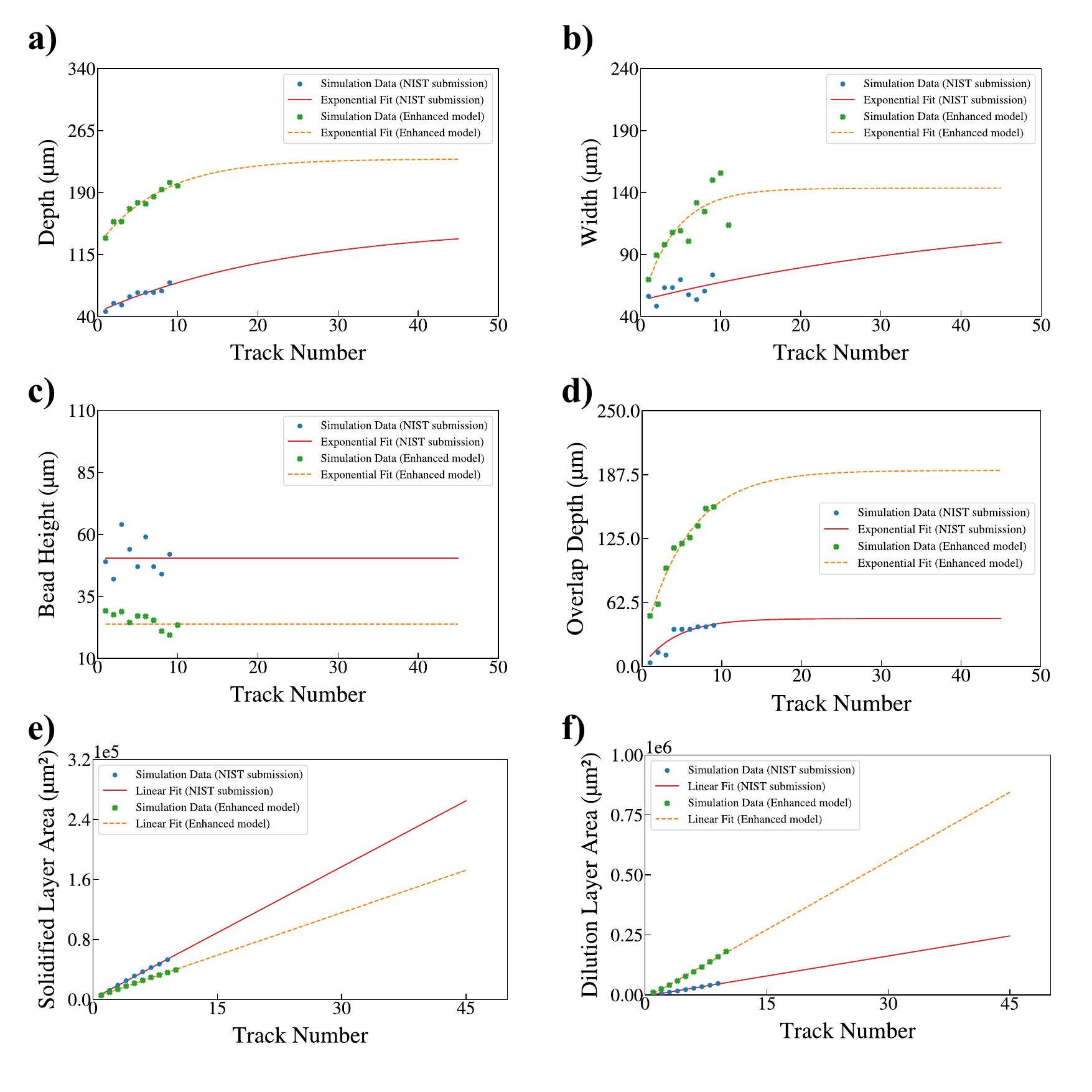}
    \caption{Powder bed of 80 \(\mu\)m prediction (5x5 mm) of 45 track average data using exponential fit for measured at x=2.545 mm}
    \label{fig:5x5_80_2.545}
\end{figure}

\begin{figure}[!h]  
    \centering
    \includegraphics[width=\textwidth]{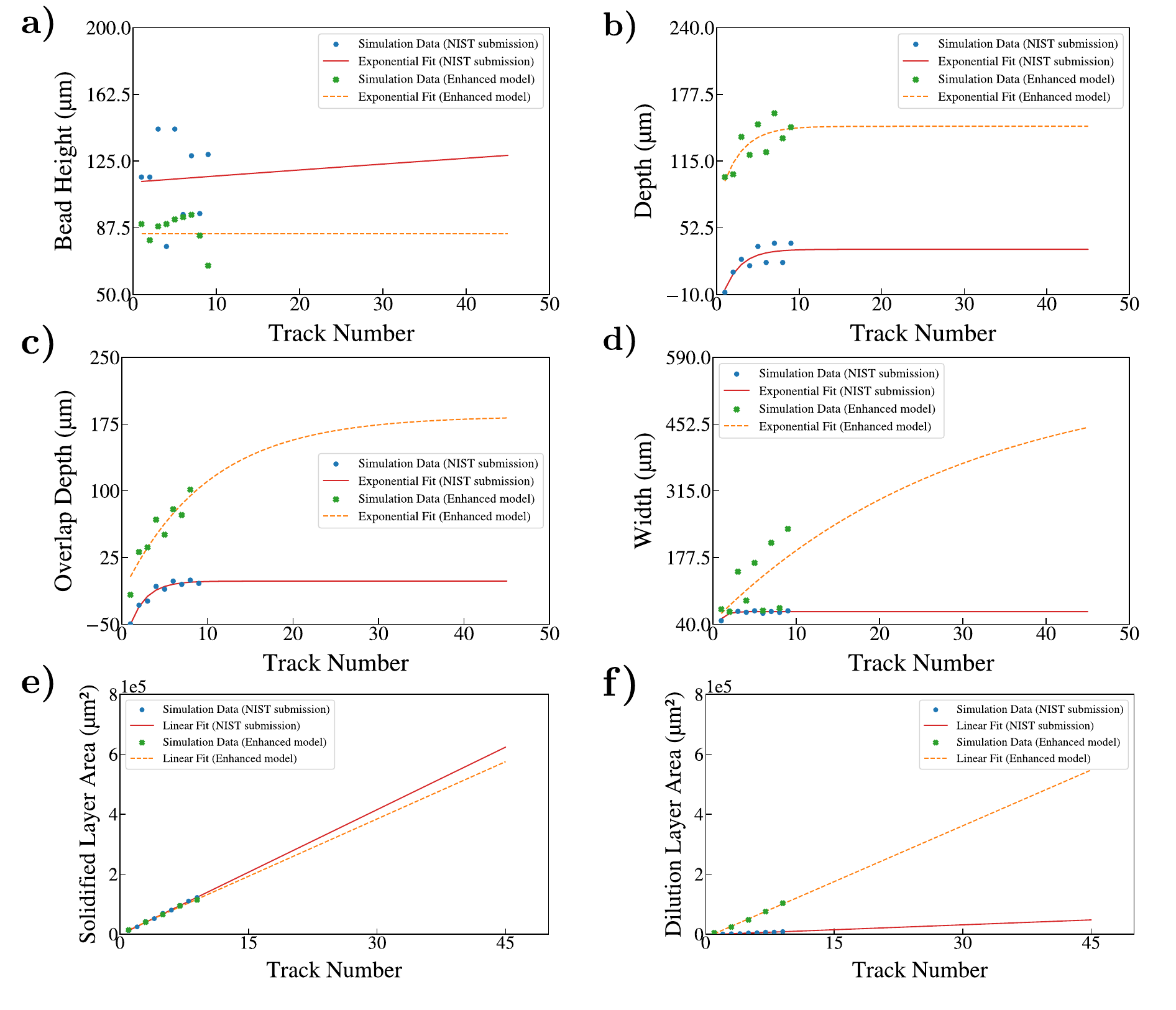}
    \caption{Powder bed of 160 \(\mu\)m prediction (5x5 mm) of 45 track average data using exponential fit for measured at x=0.460 mm}
    \label{fig:5x5_160_0.46}
\end{figure}

\begin{figure}[!h]  
    \centering
    \includegraphics[width=\textwidth]{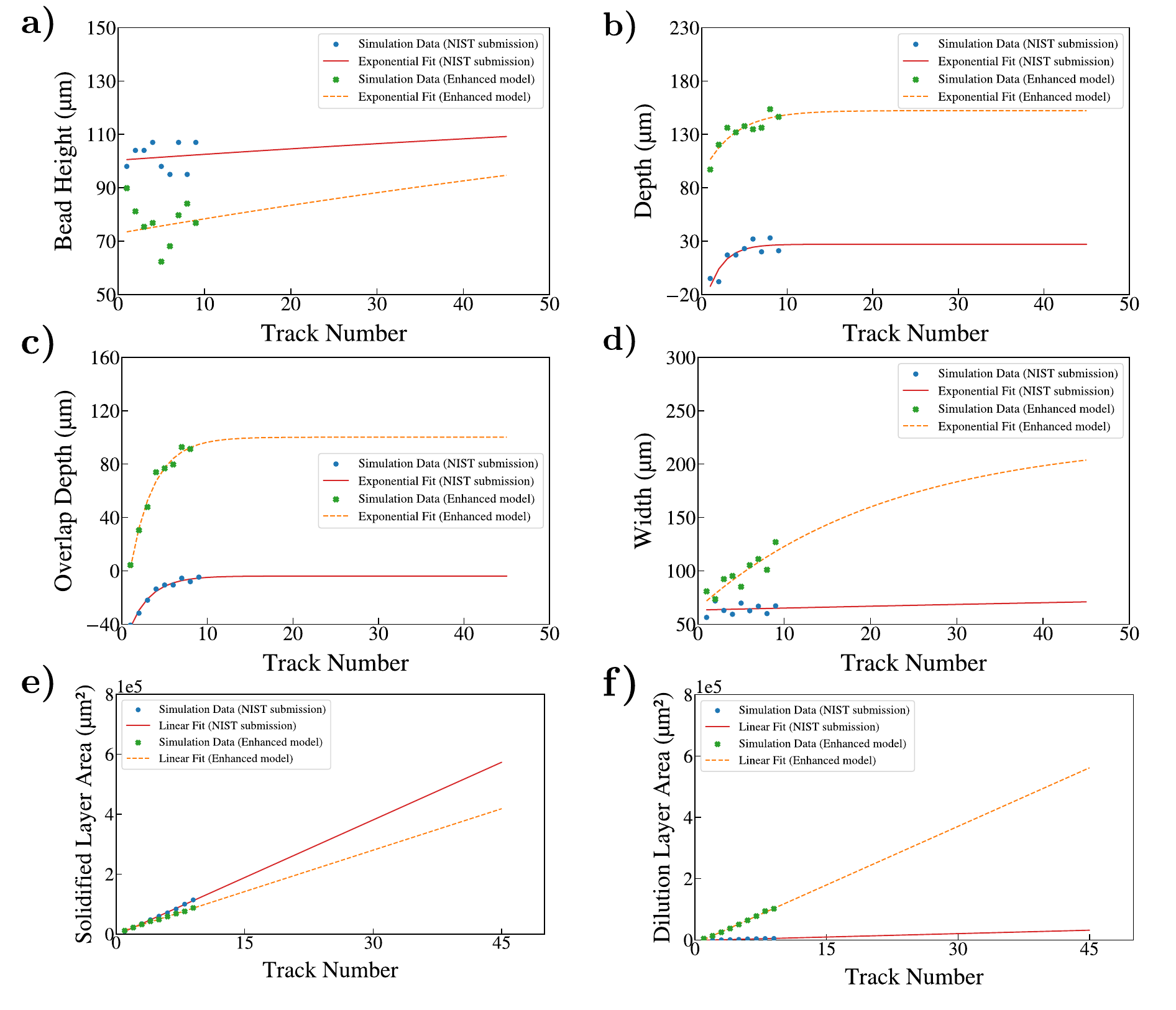}
    \caption{Powder bed of 160 \(\mu\)m prediction (5x5 mm) of 45 track average data using exponential fit for measured at x=2.545 mm}
    \label{fig:5x5_160_2.545}
\end{figure}

Due to computational limitations, up to 15 tracks were simulated for each case. To predict the data up to the 45th track, an extrapolation method was employed. Once the initial multi-track dimensions were determined, the values for depth, width, bead height, and overlap depth were fitted to an exponential saturation function of the form given in Eq.~\ref{eq:exponential}
\begin{equation}
    y(x) = A(1 - \exp(-Bx)) + C
    \label{eq:exponential}
\end{equation}
where $A$, $B$, and $C$ are fitting coefficients, and $x$ represents the track number. This fit implies a smooth transition where the dimensions undergo an initial rapid increase and eventually stabilise as $x$ increases, which supports trends observed in prior AM-Bench 2022 results \cite{amb2022results}. Conversely, for cumulative quantities such as the dilution layer area and solidified layer area, a linear fit of the form given in Eq.~\ref{eq:linear} was applied, since these metrics increase progressively and additively with each consecutive track.
\begin{equation}
    y(x) = mx + c
    \label{eq:linear}
\end{equation}
where $m$ is the slope and $c$ is the y-intercept. Figures~\ref{fig:1x5_bp}--\ref{fig:5x5_160_2.545} show the fit functions plotted against the simulated data to predict the 45-track average for all cases described in Table~\ref{tab:simulation_cases}.


\clearpage
\bibliographystyle{elsarticle-num} 
\bibliography{reference.bib}       

\end{document}